\documentclass[10pt,journal,compsoc]{IEEEtran}

\usepackage{graphicx,url}
\usepackage[utf8]{inputenc}  
\usepackage{cite}   
\usepackage{amsmath,amssymb,amsfonts}
\usepackage{times}

\usepackage[noend]{distribalgo} 
\usepackage{algorithm}  
\usepackage{pifont}
\usepackage{lipsum}  

\usepackage{textcomp}
\usepackage{float}    
\usepackage{soul}     
\usepackage{xcolor}
\usepackage{hyperref}

\newcommand{\algoritmo}[3]{\begin{algorithm}[!ht]\caption{#1}\footnotesize{\begin{distribalgo}[1] #2 #3 \end{distribalgo}}\end{algorithm}}
\newcommand{\data}[3]{\algoritmo{\textbf{- Data Types:} #1}{#2}{#3}}
\newcommand{\circuit}[3]{\algoritmo{\textbf{- ZK Circuit:} #1}{#2}{#3}}
\newcommand{\contract}[3]{\algoritmo{\textbf{- Smart Contract:} #1}{#2}{#3}}
\newcommand{\require}[2][]{\STATE $#1\textbf{require}\; #2$}
\newcommand{\events}[1][ ]{\STATE $#1\textbf{emit}\; events$}     
\newcommand{\is}[0]{\;is\;}

\newcommand{\peq}[0]{\;\mathrel{+}=\;}
\newcommand{\meq}[0]{\;\mathrel{-}=\;}
\newcommand{\get}[0]{\;\text{=}\;}
\newcommand{\eq}[0]{\;\text{==}\;}
\newcommand{\cupeq}[0]{\;\mathrel{\cup}=\;}
\newcommand{\capeq}[0]{\;\mathrel{\cap}=\;}
\newcommand{\tab}[0]{$ \ \ \ \ \ \ $}

\newcommand{\xor}[0]{\;\oplus\;}

\title{Privacy-Preserving Smart Contracts for Permissioned Blockchains: A zk-SNARK-Based Recipe Part-1}

\author{
Aldenio Burgos,
Eduardo Alchieri
\IEEEcompsocitemizethanks
   {\IEEEcompsocthanksitem Aldenio Burgos is with Banco Central do Brasil, Brazil.
    \IEEEcompsocthanksitem Eduardo Alchieri is with Universidade de Bras\' \i lia, Brazil.}
} 

\begin{document} 

\maketitle

\begin{abstract} 

The Bitcoin white paper introduced blockchain technology, enabling trustful transactions without intermediaries. 
Smart contracts emerged with Ethereum and blockchains expanded beyond cryptocurrency, applying to auctions, crowdfunding and electronic voting. 
However, blockchain's transparency raised privacy concerns and initial anonymity measures proved ineffective.
Smart contract privacy solutions employed zero-knowledge proofs, homomorphic encryption and trusted execution environments. 
These approaches have practical drawbacks, such as limited functionality, high computation times and trust on third parties requirements, being not fully decentralized. 
This work proposes a solution utilizing zk-SNARKs to provide privacy in smart contracts and blockchains. 
The solution supports both fungible and nonfungible tokens. 
Additionally, the proposal includes a new type of transactions, called delegated transactions, which enable use cases like Delivery vs Payment (DvP)

\end{abstract}

\section{Introduction}

The rise of blockchain technology, introduced by the Bitcoin white paper~\cite{nakamoto2008bitcoin}, has laid the foundation for a new era of decentralized finance, often referred to as the "Finternet"~\cite{carstens2024rise-BIS,carstens2024rise}. This envisioned financial architecture promises a globally interconnected network of systems enabling seamless transfer of value and assets. While initially focused on cryptocurrencies, blockchain's application has expanded to include smart contracts, executed through Turing-complete languages like those in Ethereum \cite{wood2014ethereum}, enabling sophisticated algorithms for applications such as auctions, crowdfunding, and electronic voting.  However, realizing the full potential of the Finternet requires addressing critical challenges, particularly in the realm of privacy.

Blockchain's transparent nature, while beneficial for auditability, raises privacy concerns.  
Its immutable public ledger exposes transaction details. 
The original Bitcoin white paper~\cite{bitcoin} proposed using pseudonymous cryptographic identifiers, not directly tied to real-world identities, to address this. 
This offers a degree of privacy by obscuring participant identities on the blockchain, though it doesn't guarantee complete anonymity.
After, various de-anonymization techniques proved this approach is not effective (e.g., \cite{biryukov2014deanonymisation, nick2015data, biryukov2019deanonymization}). 
Moreover, other popular programmable blockchains (e.g., Ethereum \cite{buterin2013ethereum}) also do not provide privacy, imposing a significant obstacle to the design of various applications where privacy is a primary requirement.

The Finternet vision, as outlined by Carstens~\cite{carstens2024rise-BIS,carstens2024rise}, hinges on security and privacy as core design principles. However, current blockchain privacy solutions often fall short of these requirements. 
Some efforts, like Zerocash/Zcash~\cite{sasson2014zerocash,zcash} and Monero \cite{noether2016ring}, primarily address privacy in the context of cryptocurrencies.  
Others focus on smart contracts, employing techniques like zero-knowledge proofs (e.g., \cite{xiong2023verizexe,jiang2024rpsc}), homomorphic encryption (e.g., \cite{diamond2021anonzether,steffen2022zeestar}), delegation to trusted-execution environments (e.g., \cite{secretnetworkgraypaper,luo2024ethercloak}), or trusted third parties (e.g., \cite{kosba2016hawk,kalodner2018arbitrum}). 
However, these approaches often present significant drawbacks in the context of a fully decentralized and interoperable Finternet. 
For instance, Anonymous Zether~\cite{diamond2021anonzether} suffers from limitations like restricted token types and account freezes. 
Homomorphic encryption introduces substantial computational overhead, hindering scalability, while delegation to trusted environments compromises the decentralized nature of the Finternet.

To realize the full potential of the Finternet, a new approach to privacy is needed—one that is scalable, flexible, and compatible with a decentralized, multi-chain environment. This work proposes a novel solution using zk-SNARKs (Zero-Knowledge Succinct Non-Interactive Argument of Knowledge) to achieve privacy in smart contracts and blockchain transactions within permissioned networks, a likely foundation for many Finternet applications. Our solution directly addresses the limitations of existing approaches by supporting both fungible and nonfungible tokens, leveraging a UTXO model for enhanced parallelism, and introducing a novel concept of delegated transactions. These features enable complex, privacy-preserving interactions crucial for Finternet use cases, such as atomic Delivery vs Payment (DvP) settlements, which we will demonstrate in this paper. Our approach offers a recipe for building a truly private and secure foundation for the future of finance.

The remainder of this text is organized as follows. Section~\ref{sec:basic} presents some basic concepts used in the proposed solutions. Section~\ref{sec:proposal} presents the proposal for privacy in smart contracts and blockchains, while Section~\ref{sec:usecase} discusses a DvP use case. Finally, Section~\ref{sec:conclusion} concludes the paper.

\section{Basic Concepts}
\label{sec:basic}

This section introduces some important basic concepts used in our proposal.

\subsection{Merkle Trees}

Merkle trees are fundamental data structures in cryptography and computer science, widely used to ensure data integrity and efficient verification. 
They play a crucial role in various applications, including blockchain technology, distributed version control systems, and digital signatures.  
Algorithm~\ref{alg:merkle} presents the $getRoot$ function, which calculates the root hash of a Merkle tree. 
This function enables the verification of data inclusion in the tree without requiring access to the entire dataset.

The $getRoot$ algorithm takes two inputs: a $value$ representing the data element to be verified, and a $path$, which is a Merkle path data structure ($MPat$). 
The $MPat$ encodes the position of the $value$ within the Merkle tree as a list of $MerkleStep$ elements. 
Each $MerkleStep$ contains the hash of the sibling node and a boolean value indicating whether the current node is on the left or right side of its parent.

The algorithm begins by initializing a variable $h$ with the input $value$.  
It then iterates through each $MerkleStep$ in the provided $path$. 
For each step, the algorithm performs a hash computation. 
If the current node is on the left side of its parent, it concatenates the sibling's hash with the current value of $h$ and computes their combined hash using the $hash256$ function. 
Otherwise, it concatenates $h$ with the sibling's hash and computes their combined hash. 
The result of this computation is then assigned back to $h$. 
This process is repeated for each step in the Merkle path.

Finally, after processing all steps, the algorithm returns the final value of $h$, which represents the calculated Merkle root. 
By traversing the tree from the input $value$ up to the root and performing the specified hash computations, the algorithm effectively reconstructs the root hash. 
This allows for verification of the data's inclusion in the tree without needing to access or process the entire dataset.

        
\algoritmo{Merkle Tree}{\label{alg:merkle}}{ 
\INDENT{\textbf{Data Types:}}
    \STATE $MRoot: uint256$                             \COMMENT{Merkle tree root}                     \label{data:merkle.root}
    \STATE $MPat: List\langle MerkleStep \rangle$        \COMMENT{Merkle path}                          \label{data:merkle.path}
    \STATE $MerkleStep:$                                 \COMMENT{Merkle path step}                     \label{data:merkle_step} 
        \STATE \tab $hash: uint256,$                     \COMMENT{hash of the child nodes}              \label{data:merkle.step.hash}
        \STATE \tab $at\_left: bool$                     \COMMENT{position in the tree: left or not}    \label{data:merkle.step.direction}
\ENDINDENT

\vspace{2.5mm}
\INDENT{$\textbf{getRoot}(value: uint256, path: MPat) : uint256$}                       \label{cir:merkle_root}  
    \STATE $h \get value$    
    \FORALL{$step \in path$}
        \IF {step.at\_left}
            \STATE $h \get hash256(step.hash \cup h)$
        \ELSE
            \STATE $h \get hash256(h \cup step.hash)$
        \ENDIF
    \ENDFOR    
    \RETURN $h$       
\ENDINDENT
}

\subsection{Zero Knowledge Cryptography} \label{sec:zk}

Zero-Knowledge Proofs (ZKPs) are a revolutionary cryptographic technique enabling one party to prove the validity of a statement without revealing any information (zero knowledge). 
This method ensures confidentiality and authenticity, safeguarding sensitive data.
ZKPs possess three essential properties: completeness, soundness and zero-knowledge.
Completeness guarantees the verifier accepts true statements, while soundness ensures false statements are rejected.
Zero-knowledge ensures the verifier learns nothing beyond the statement's validity. 
These properties foster trustless interactions, mitigating risks associated with data exposure.  
ZKPs empower innovative solutions by balancing privacy and verification.

Zero-Knowledge Succinct Non-Interactive Argument of Knowledge (zk-SNARKs) is a type of zero-knowledge proof that enables efficient and scalable verification of complex computations. 
Zk-SNARKs allow a prover to demonstrate the validity of a statement without revealing private data, ensuring privacy and security. 
Zk-snarks is characterized by its succinctness, efficiency and non-interactivity, it provides fast verification times and minimal computational overhead. 
Additionally, a ZK circuit is a specific tool or technique used to encode computer programs as ZK proofs, defining the rules and logic to verify that a program was run correctly.

Our proposal uses ZK circuits (based on zk-SNARKs) to prove that participants knows some data, without revealing it. 
In summary, firstly two parameters $S_p$ and $S_v$ are created to the prover and verifier, respectively. 
At the prover, a ZK circuit receives a public statement $x$, a private witness $w$ and uses $S_p$ to produce a proof $\pi$. 
After, a verifier uses the proof $\pi$, the public statement $x$, and the parameter $S_v$ to accept or reject that proof, i.e., to accept/reject that the prover posses the private data $w$. 
Notice the verifier does not access $w$. 
In the algorithms presented in this paper, we use the hypothetical symbolic function $convertToProof(w)$ to abstract the creation of $\pi$ at the prover, once that $\pi$ is used by the verifier to check that the correct witness was provided to the circuit.

\subsection{Tokens}
The term ``token" refers to a digital representation of traditional assets, created and recorded on a programmable platform. 
In essence, a token is a unit of value or asset that is stored in a digital system and can represent various goods, such as currencies, stocks, real estate, among others. 
This representation is maintained in databases using technologies such as Distributed Ledger Technology (DLT), which facilitates the updating of a shared ledger, allowing the execution of transactions such as issuance, trading and settlement of financial assets. 
The digital and programmable nature of tokens enables the development of complex financial functions and the implementation of efficient and secure systems for the movement of assets.
In the context of tokenization of currencies and financial assets, tokens evolve in different stages. In the first stage, they are used as a mere digital representation of the value of an asset. As they move to the second stage, they incorporate business functionalities, allowing for more sophisticated financial operations. In the third stage, tokens become highly composable, allowing for the modular creation of new financial products and services. This process of evolution and flexibility of tokens is at the base of the development of more efficient, inclusive and accessible financial systems, as exemplified by Central Bank Digital Currency (CBDC) projects.
Burning a nonfungible token (NFT) means permanently removing it from circulation. This is usually done by sending the NFT to an inaccessible wallet address, effectively destroying it.

\section{Proposed Solution} \label{sec:proposal}

The proposed system is composed of:
\begin{itemize}
    \item the network's managing institution, i.e., the network authority (e.g., the Central Bank of Brazil).
    \item a set of participating institutions, duly authorized by the network's managing institution.
    \item a private and off-chain communication network between participants.
    \item a vanilla permissioned programmable blockchain network, composed of at least one node per participating institution.
    \item a set of UTXO-based smart contracts for token management, the $TK$ contracts.
    \item an off-chain system for constructing zero-knowledge proofs, with various circuits; this system will be executed privately by each participant.
    \item a set of zero-knowledge proof verification contracts; each circuit in use must have its verifier published on the blockchain.
    \item a set of business smart contracts that will use all this structure to deliver sophisticated services.
\end{itemize}

Each transaction submitted to the blockchain is composed of public data and zero-knowledge proofs, when necessary. 
In fully public transactions, the blockchain would function normally. 
In transactions that involve private data, the guarantees and validations on the private data are provided by the zero-knowledge circuits. 
Considering the use of UTXO-based contracts, several relationships (input $\rightarrow$ output) can be validated in secrecy, while the guarantee of double-spending and other checks can be performed clearly by the contract, using the commitments, nullifiers, grabbers and the available public data. 
The output tokens are transferred to their respective owners directly without compromising the privacy or security of the system.

The implementation proposed here offers a set of significant advantages and opportunities, reflecting its potential impact and utility in various application scenarios. It is built upon a foundation of key principles designed to ensure its robustness, security, and efficiency. 
These principles can be categorized into aspects related to the blockchain and execution environment, transparency and auditability, decentralization and resilience, and finally, performance and scalability.

Regarding the blockchain and execution environment, the solution prioritizes atomicity, platform agnosticism, and consensus independence.
Transactions are guaranteed to be atomic, ensuring they execute completely or not at all, thanks to the inherent properties of programmable blockchain platforms. Furthermore, while initially designed for Ethereum Virtual Machine (EVM) compatible blockchains, the solution can be deployed on any smart contract platform.  It also functions seamlessly regardless of the underlying consensus mechanism used by the blockchain.

Transparency and auditability are addressed through selective transparency and built in auditability features. 
The solution offers granular control over data visibility, allowing transparency and privacy to be adjusted at the field or attribute level without sacrificing overall programmability. 
Additionally, the system is designed to allow authorized entities to audit transactional data, with a customizable level of detail exposed during an audit.

Decentralization and resilience are achieved by ensuring there is no single point of failure and by promoting a fully decentralized architecture.
All nodes in the network operate identically, eliminating any single point of failure. 
The uniform role of all network nodes further ensures maximum decentralization.

Finally, concerning performance and scalability, the solution leverages parallel processing and offers high performance even with computationally intensive operations.  
It utilizes a UTXO (Unspent Transaction Output) model, similar to Bitcoin, to achieve a high degree of parallelism, further enhanced by off-chain transaction construction. 
Each participant can create multiple transactions simultaneously, limited only by their processing power and the degree to which their state is fractionated. 
While there is a computational cost in constructing Zero-Knowledge (ZK) proofs, this is mitigated by off-chain parallelism. 
Moreover, with the adoption of ZK-SNARKs, the computational cost added to the blockchain for proof verification is logarithmically related to proof construction, rendering it negligible in the overall performance of the blockchain.
Importantly, the proposed solution is well-behaved and does not interfere with the operation of any other solution deployed on the same blockchain network.

\subsection{Choice of zk-SNARKs}
\label{sec:zk-SNARKs-choice}

Our solution employs zk-SNARKs (Zero-Knowledge Succinct Non-Interactive Arguments of Knowledge) as the underlying zero-knowledge proof system. 
This choice is motivated by several key advantages that zk-SNARKs offer in the context of blockchain privacy and our specific design goals.  
zk-SNARKs generate proofs that are very small in size, typically a few hundred bytes, regardless of the complexity of the computation being proved~\cite{groth2016size}.
This succinctness is crucial for blockchain applications, where storage space is a premium. 
Small proofs minimize the on-chain footprint of our privacy solution, contributing to scalability. 
Furthermore, zk-SNARKs enable extremely fast proof verification, typically in the order of milliseconds~\cite{groth2016size}. 
This efficiency is essential for maintaining the performance of the blockchain network, as nodes can quickly verify the validity of transactions without significant computational overhead. 
Another crucial aspect is their non-interactivity: zk-SNARKs are non-interactive, meaning the prover can generate a proof without any back-and-forth communication with the verifier~\cite{goldwasser1989knowledge}. 
This property is crucial for asynchronous blockchain environments where transactions are typically submitted and verified in separate steps. 
Finally, the ecosystem around zk-SNARKs has matured significantly in recent years, with the development of various libraries and tools (e.g., libsnark~\cite{libsnark}, ZoKrates~\cite{zokrates}, Circom~\cite{circom}) that simplify the process of creating and verifying zk-SNARK proofs. 
This allows us to leverage existing expertise and infrastructure.

While other zero-knowledge proof systems like zk-STARKs and Bulletproofs offer certain advantages, they are currently less suitable for our specific needs. 
zk-STARKs (Zero-Knowledge Scalable Transparent Arguments of Knowledge) provide post-quantum security and do not require a trusted setup, unlike most current zk-SNARK constructions~\cite{ben2018scalable}. 
However, they produce larger proofs (several kilobytes to megabytes) and have slower verification times compared to zk-SNARKs~\cite{bunz2020transparent}. 
These factors make them less practical for on-chain verification in a high-throughput blockchain environment. 
Bulletproofs are another promising ZKP system that does not require a trusted setup and has relatively small proof sizes~\cite{bunz2018bulletproofs}. 
However, their verification time scales linearly with the size of the proof computation, making them less efficient than zk-SNARKs for complex ZK circuits. 
Moreover, Bulletproofs are particularly well-suited for range proofs, which are not our primary focus.

It is important to acknowledge that most current zk-SNARK constructions rely on a trusted setup phase to generate public parameters~\cite{groth2016size}. 
This phase requires careful execution to prevent the creation of a ``toxic waste'' that could compromise the security of the system. 
While this is a potential drawback, various techniques, such as multi-party computation (MPC) ceremonies, have been developed to mitigate the risks associated with trusted setups~\cite{bayer2023efficient}. 
Furthermore, ongoing research is exploring zk-SNARK constructions that eliminate or minimize the need for a trusted setup (e.g., ``universal'' or ``transparent'' SNARKs)~\cite{kosba2019aurora}. 
Our solution could potentially transition to such constructions as they become more mature and practical. 
In conclusion, zk-SNARKs offer the best balance of succinctness, verification efficiency, and non-interactivity for our privacy-preserving smart contract solution. 
While we acknowledge the trusted setup and post-quantum security requirement, we believe that the benefits of zk-SNARKs outweigh the drawbacks in the context of our design goals and that ongoing research will continue to address these limitations.

In our algorithms (e.g., Algorithms~\ref{cir:mint}, \ref{cir:trans}, \ref{cir:rev}, \ref{cir:hid}, \ref{cir:grab}, \ref{cir:del_mint}, \ref{cir:del_trans}, \ref{cir:del_rev}, \ref{cir:del_hid}, \ref{cir:dvp}), we use the symbolic function $convertToProof(wit)$ to represent the process of generating a zk-SNARK proof from a given witness wit. 
Conceptually, this function encapsulates the complex cryptographic operations involved in zk-SNARK proof generation. 
It takes the witness data, which contains private information, and encodes it into a suitable format for the chosen zk-SNARK system. 
Then, it executes the core proving algorithm to produce a succinct proof, denoted as $\pi$.
This proof attests to the fact that the prover possesses a valid witness that satisfies the constraints defined in the corresponding ZK circuit, without revealing the witness itself.
The generated proof $\pi$ can be efficiently verified using the public inputs and the verification key.
The specific implementation of $convertToProof$ would depend on the chosen zk-SNARK library (e.g., libsnark~\cite{libsnark}, ZoKrates~\cite{zokrates}) and the underlying proving system (e.g., Groth16~\cite{groth2016size}, PLONK~\cite{gabizon2019plonk}).

\subsection{Tokens}

Algorithm~\ref{alg:token} defines the data structures of types related to a token.
The $TPre$, or Token preimage, at line~\ref{data:token_pre_image} represents the preimage of a single token. 
This structure contains information such as the owner's account ($owner$), the token type ($type$), a giant random number called $nonce$ (derived from '\textit{nonsense}') to ensure the uniqueness of the token, its quantity ($amount$) if it represents a fungible entity, or its $id$ if it represents a nonfungible entity. 
Optionally, a payload that allows new fields to be included as needed.
For example: the fungible token owner's taxpayer number or the title and description of a rare and exclusive bottle of wine in the case of an NFT. 
Both the $owner$ and the $nonce$ are 256-bit unsigned integers, meaning they can have a value between 0 and ($2^{256}-1$) \footnote{{115,792,089,237,316,195,423,570,985,008,687,907,853,269,984,665, 640,564,039,457,584,007,913,129,639,935}}. 
This choice is due to the standard EVM word size and the security level we can reach with 256 bits.
There is nothing that prevents these fields from having another type on a different platform.

The $type$, $amount$ and $id$ fields are depicted as 256-bit unsigned integers either for educational purposes.
In practice, these fields can be much smaller for performance and memory reasons.
We understand that a token will hardly be fungible and nonfungible at the same time.
However, throughout this article, we will work as if both fields were always present; 
in a fungible token, the $id$ will be zero, and in a nonfungible token, the $amount$ will be zero.

The payload field is an optional space that can carry any data structure, of any type or size, regarding the Token, that is needed for a specific use case.

\data{token}{\label{alg:token}}{
\STATE $Account: uint256$                       \COMMENT{token owner account}                       \label{data:token_Account_type}   
\STATE $Nonce: uint256$                         \COMMENT{nonce type}                                \label{data:nonce_type} 
\STATE $ID: uint256$                            \COMMENT{ID type}                                   \label{data:id_type} 
\STATE $Amount: uint256$                        \COMMENT{amount type}                               \label{data:amount_type} 
\STATE $Type: uint256$                          \COMMENT{token type}                                \label{data:type_type} 
\STATE $TPre: $                                 \COMMENT{token preimage}                           \label{data:token_pre_image} 
    \STATE \tab $owner: Account,$               \COMMENT{owner account}                             \label{data:token_pre.owner}
    \STATE \tab $type: Type,$                   \COMMENT{token type}                                \label{data:token_pre.type}
    \STATE \tab $nonce: Nonce,$                 \COMMENT{entropy adder}                             \label{data:token_pre.nonce}
    \STATE \tab $amount: Amount,$               \COMMENT{fungible token amount}                     \label{data:token_pre.amount}
    \STATE \tab $id: ID,$                       \COMMENT{nonfungible token identifier}             \label{data:token_pre.id}
    \STATE \tab $[payload: any]$                \COMMENT{optional, any useful information}          \label{data:token_pre.payload} 
}

\subsubsection{Commitment} \label{sec:token_commitment}

In our system, a \textbf{token commitment} is a cryptographically secure representation of a token that hides the token's details while allowing for verification of its existence and properties. Thanks to the hash function properties, the token commitment does not allow any inference regarding the content of its preimage. 
Formally, let $H$ be a cryptographic hash function, $H: \{0,1\}^* \rightarrow \{0,1\}^n$, where $n$ is the output length in bits (e.g., 256 for SHA-256~\cite{sha256}). Let $t$ be a token preimage as defined in Algorithm~\ref{alg:token}, which includes a unique nonce. 
A commitment $C$ and a partial commitment $PC$ to the token $t$ are defined as:
\[ C(t) = H(\texttt{PC}(t) \parallel t.\texttt{payload}) \]
where:
\[ \texttt{PC}(t) = H(t.\texttt{owner} \parallel t.\texttt{type} \parallel t.\texttt{amount} \parallel t.\texttt{id} \parallel t.\texttt{nonce}) \]

The symbol $\parallel$ denotes concatenation, and $t.payload$ is the optional payload of the token. 
If the payload is not present, then the commitment is simply $C(t) = PC(t)$. 
The ZK circuit in Algorithm~\ref{cir:token_commitment} implements this commitment scheme, ensuring that the commitment can be verified without revealing the underlying token preimage $t$.

On line~\ref{data:token_commitment}, we have the $TCom$ type that represents a \textbf{token commitment}. 
The ZK circuit function $commit$ (line~\ref{cir.commit}) calculates the cryptographic commitment of the token preimage $TPre$ (Algorithm~\ref{alg:token}), while the $partialCommit$ function (line~\ref{cir.partial_commit}), in turn, hashes the main fields of the token, returning a single value that represents them.

\circuit{Create Commitment}{\label{cir:token_commitment} }{
\INDENT{\textbf{Data Types:}}
    \STATE $TCom: uint256$    \COMMENT{token commitment type}  \label{data:token_commitment} 
\ENDINDENT
\vspace{2.5mm}
\INDENT{\textbf{commit}$(t: TPre) : TCom$}     \label{cir.commit}
    \STATE $h \get partialCommit(t)$                                    
    \RETURN $(t.payload \eq \bot)?\;h: hash256(h \parallel t.payload)$   \label{cir.commit_return}
\ENDINDENT
\vspace{2.5mm}
\INDENT{\textbf{partialCommit}$(t: TPre) : uint256$}        \label{cir.partial_commit}
    \RETURN $ hash256(t.owner \parallel t.type  \parallel t.amount \parallel t.id \parallel t.nonce)$        
\ENDINDENT
}

\subsubsection{Nullifier}

A \textbf{token nullifier} is a unique identifier derived from a token that is used to prevent double-spending. 
It is responsible for consuming, or nullifying, a valid token, without leaving room for deanonymization through transaction graph analysis (\cite{biryukov2014deanonymisation}, \cite{ron2013quantitative}, \cite{nick2015data} and \cite{biryukov2019deanonymization}). 
Once a token is spent, its corresponding nullifier is published, rendering any further attempts to spend the same token invalid. 
The nullifier is constructed in a way that reveals no information about the original token, preserving privacy. 
Formally, let $C$ be the commitment function as defined in Section~\ref{sec:token_commitment}. 
Let $t$ be a token preimage as defined in Algorithm~\ref{alg:token}, and $sk$ be the secret key of the token's owner. 
A nullifier $N$ for token $t$ is defined as:
\[ N(t, sk) = C(t') \]
where $t'$ is a modified version of $t$ such that the $owner$ field of $t$ is replaced with the owner's secret key.

The zk circuit, in Algorithm~\ref{cir:token_nullifier}, ensures that the construction of the nullifier from the token preimage data is correct. 
This process guarantees that the nullifier is uniquely tied to the token and the owner's secret key while revealing no information about either.
The advantage of replacing the token commitment with its nullifier when trying to consume it lies in the impossibility of inferring any useful information about the transaction data, including the relationship between consumed tokens and created tokens, from the data shared on the blockchain. 

The $TNul$ type (line~\ref{data:token_nullifier_type}) represents a \textbf{token nullifier}. 
The main function, called $nullify$ (line~\ref{cir.nullify}), which receives two parameters: an instance of Token preimage $TPre$, represented by $t$, and the secret key $secret\_key$ of the token owner\footnote{Only the owner of a token can create the nullifier for that token}. 
In addition, the auxiliary function $getAccount$ (line~\ref{cir.get_account}) receives a secret key $secret\_key$ as a parameter and returns the associated account number, of type $Account$ (line~\ref{data:token_Account_type}).

The first step in the process is to derive the account associated with the provided secret key, through the $getAccount$ function. 
Then, a check is performed with the $require$ statement, which ensures equality between the generated account and the value of the $owner$ attribute of the token. 
This check ensures that the provided secret key actually corresponds to the owner of the token, ensuring that only the legitimate holder of the token can nullify it. 
After validation, the value of the $owner$ attribute of the token is replaced by the secret key $secret\_key$ of its owner.
Finally, the token nullifier is generated by calling the $commit$ function executed on the newly modified token.

The $getAccount$ function (line~\ref{cir.get_account}) calculates the account number linked to a secret key.
In its first step, it derives a public key $public\_key$ linked to the provided secret key $sk$, through the $derivePublicKey$ function. 
This is a basic cryptographic function and is present in the main ZK libraries, but may have other names. 
Next, the hash of the $public\_key$ is calculated and returned, which is then used to verify the owner of the token.

\circuit{token nullifier}{\label{cir:token_nullifier}}{
\INDENT{\textbf{Data Types:}}
    \STATE $TNul: uint256$                      \COMMENT{token nullifier}       \label{data:token_nullifier_type}
    \STATE $Bytes: Array\langle uint \rangle$   \COMMENT{Byte Array}            \label{data:byte_array_type}
    \STATE $SKey: uint256$                      \COMMENT{Secret key type}       \label{data:secret_key_type}    
    \STATE $PKey: Bytes$                        \COMMENT{Public key type}       \label{data:public_key_type}
\ENDINDENT
\vspace{2.5mm}
\INDENT[creates a nullifier]{\textbf{nullify}$(t: TPre, sk: SKey): TNul$} \label{cir.nullify}
    \require $t.owner \eq getAccount(sk)$       \COMMENT{confirms ownership}
    \STATE   $t.owner \get sk$                  \COMMENT{replaces owner with secret key}
    \RETURN  $commit(t)$                        \COMMENT{returns a commitment -- alg.~\ref{cir:token_commitment}}
\ENDINDENT
\vspace{2.5mm}
\INDENT[gets account from secret key]{\textbf{getAccount}$(sk: SKey) : Account$}   \label{cir.get_account}                                               
    \STATE   $public\_key \get derivePublicKey(sk)$  \COMMENT{derives the public key}
    \RETURN  $hash256(public\_key)$                  \COMMENT{returns the hash of the public key}
\ENDINDENT
}

\subsubsection{Grabber}

A \textbf{token grabber}, or grabber, is a cryptographic commitment that allows a designated authority to seize a token under specific circumstances, such as complying with a court order. 
The grabber is constructed using a special key associated with the token's owner, the grabber key.  
When participants joins the network, they must generate their Grabber Keys, one for each token contract $TK$ available in the system. 
This key is generated by the participant, using its secret key and the contract's public grabber nonce. 
The participant's grabber key for each contract must be securely shared with the token's contract authority, which will validate them during the participant's authorization process, ensuring that only the authority or the participant can generate a token grabber for this participant's tokens.
If, later, a new token contract is added to the system, each participant who wishes to operate with that contract must create their respective grabber key and send it to the contract authority.

Formally, let $C$ be the commitment function as defined in Section~\ref{sec:token_commitment}. Let $t$ be a token preimage as defined in Algorithm~\ref{alg:token}, and $gk$ be a grabber key associated with the token's owner. The grabber key is generated as $gk = \text{cipher}_{sk}(\text{nonce}_g)$, where $sk$ is the owner's secret key, $\text{cipher}$ is a symmetric encryption function, and $\text{nonce}_g$ is a unique nonce associated with the token contract. The grabber $G$ for token $t$ is defined as:
\[ G(t, gk) = C(t'') \]
where $t''$ is a modified version of $t$ such that the \texttt{owner} field of $t$ is replaced with the correct grabber key $gk$.

Algorithm~\ref{cir:token_grabber} describes the data types and zk circuits that allows the creation of a token grabber. 
It starts defining two data types used within the Token smart contract. 
The $TGra$ data type used to identify a token grabber and the $GKey$, which represents the grabber key type. 
The participants use the function $createGrabberKey$ (line~\ref{cir.create_grab_key}), with their secret key $sk$ and the grabber nonce associated with a contract as parameters (each contract has a unique grabber nonce, as we will see later), to create their grabber key for that contract. 
Verification of the validity of a grabber key is simple, just decrypt it with the participant's public key and check if the result is the contract's nonce.

The $grab$ function (lines~\ref{cir.grab}--\ref{cir.grab.return}) receives as input the preimage $t$ of a token and the corresponding participant's grabber key $gk$ to generate a grabber. 
For this, the token owner is changed to the grabber key $gk$ and the commitment of the modified token is calculated, thus generating the $TokenGrabber$. 
Note that only the authority and the token owner themselves have $gk$, being the only participants who can generate this grabber token. 
As we will see later, the authority generates a grabber to grab the corresponding token $t$, while the owner of $t$ also generates the corresponding grabber to prove, when consuming $t$, that $t$ has not yet been grabbed.

\circuit{token grab}{\label{cir:token_grabber}}{
\INDENT{\textbf{Data Types:}} 
    \STATE $TGra: uint256$              \COMMENT{token grabber}     \label{data:token_grabber} 
    \STATE $GKey: uint256$              \COMMENT{grabber key}       \label{data:grabber_key} 
\ENDINDENT
\vspace{2.5mm}
\INDENT{\textbf{createGrabberKey}$(sk: SKey, nonce_g: Nonce) : GKey$}       \label{cir.create_grab_key}       
    \RETURN $sk.cipher(nonce_g)$ \COMMENT{cipher nonce with sk}             \label{cir.create_grab_key.return}       
\ENDINDENT
\vspace{2.5mm}
\INDENT{\textbf{grab}$(t: TPre, gk: GKey) : TGra$}  \label{cir.grab}
    \STATE   $t.owner \get gk$      \COMMENT{change owner by grabber key}
    \RETURN  $commit(t)$            \COMMENT{return a commitment -- alg.~\ref{cir:token_commitment}} \label{cir.grab.return}
\ENDINDENT
}


\subsection{The Token's Smart Contract}

This section presents the token based smart contract which together with the ZK circuits create the bedrock of this privacy proposal.

First we present the smart contract overview, its interface and state variables, then the protocols for executing several basic flows, such as:
\begin{itemize}
    \item Mint: Issues new tokens from an asset. For example, this transaction can be used to create tokens from money deposited with an authority like the Central Bank.
    \item Transfer/Burn: Used to transfer assets represented by tokens by consuming tokens and generating new ones or burning them.
    \item Revealing transfer: Used to reveal the content of a token, for example, by consuming a fungible token and adding the value of this token to an account.
    \item Hiding transfer: Used to hide a token, for example, by removing the value from an account and creating a token with private data.
    \item Grabbing: Used by an authority to grab a token.
\end{itemize}

After, we also present delegated versions for these transactions (except for grabbing since it does not make sense to have a delegated grab). 
The general idea behind these transaction is that the token owner can delegate the permission to execute a transaction with the related token to other parties. 
As we will see later, these transactions allow more evolving use cases (e.g., DvP-Delivery versus Payment).

\subsubsection{Overview}

Here we present the token contract main functions and state variables.
The Token smart contract utilizes a variety of state variables to manage its operations.  
$type\_t$ stores the token type as a 128-bit unsigned integer.  
A Merkle tree, $tree\_c$, is employed to store commitments, likely for token-related data.  
The authority's Ethereum's external owned address (EOA) is stored in $auth\_add$, while their account identifier is kept in $auth\_acc$.  
An audit account, $audit\_acc$, is also maintained.  
Authorized issuers are tracked in the $issuers$ mapping, with a boolean value indicating their status.  
Another Merkle tree, $tree\_i$, appears to be related to issuer management, potentially storing commitments to their identities or permissions. 

To prevent double-spending, $nullifiers$ are recorded in a mapping.  
Similarly, $grabbers$, potentially entities with special roles, are tracked in the $grabbers$ mapping.  
$grab\_nonce$ provides entropy for grabber-related operations.

\contract{Token (TK)}{\label{alg:token_smart_contract}}{
\INDENT{\textbf{Data Types:}}  
    \STATE $FBal: Map\langle uint256, uint256 \rangle$
    \STATE $NFBal: Map\langle uint256, List\langle uint256 \rangle\rangle$
\ENDINDENT
\vspace{2.5mm}    
\INDENT{\textbf{State Variables:}}  
    \STATE $ type_t: uint256$                               \COMMENT{contract's token type}         \label{var:token.token_type}         
    \STATE $ tree_c: MerkleTree\langle TCom \rangle$        \COMMENT{commitment tree}               \label{var:token.commitments_tree}   
    \STATE $ auth\_add: address$                            \COMMENT{authority's EOA}               \label{var:token.auth_eoa}           
    \STATE $ auth\_acc: Account$                            \COMMENT{authority's account}           \label{var:token.auth_acc}          
    \STATE $ audit\_acc: Account$                           \COMMENT{audit account}                 \label{var:token.audit_acc}          
    \STATE $ issuers: Map\langle address, bool \rangle$     \COMMENT{public issuers' address}       \label{var:token.issuers}         
    \STATE $ tree_i: MekleTree\langle Account \rangle$      \COMMENT{hidden issuer tree}            \label{var:token.issuer_tree}        
    \STATE $ nullifiers: Map\langle TNul, bool\rangle$      \COMMENT{known nullifiers}              \label{var:token.nullifiers}       
    \STATE $ grabbers: Map\langle TGra, bool\rangle$        \COMMENT{known grabbers}                \label{var:token.grabbers}          
    \STATE $ grab\_nonce: Nonce$                            \COMMENT{grabber entropy}               \label{var:token.grab_nounce}      
    \STATE $ balances: FBal$                                \COMMENT{open balances}                 \label{var:token.bal}
    \STATE $ nfts: NFBal$                                   \COMMENT{open NFTs}                     \label{var:token.nfts}
    \STATE{ $\textit{\textbackslash*}\; verifiers'\; smart\;contracts\; \textit{*\textbackslash}$} 
    \STATE $ mint_{ver}: MV $                   \COMMENT{mint verifier}                     \label{var:token.mint_verifier}
    \STATE $ del\_mint_{ver}: DMV$              \COMMENT{delegated $^{\prime\prime}$}       \label{var:token.del_mint_ver}
    \STATE $ transf_{ver}: TV$                  \COMMENT{transfer verifier}                 \label{var:token.trans_verifier}
    \STATE $ del\_transf_{ver}: DTV$            \COMMENT{delegated $^{\prime\prime}$}       \label{var:token.del_trans_ver}
    \STATE $ rev\_transf_{ver}: RTV$            \COMMENT{revealing transfer verifier}       \label{var:token.rev_trans_ver}
    \STATE $ del\_rev\_transf_{ver}: DRTV$      \COMMENT{delegated $^{\prime\prime}$}       \label{var:token.del_rev_trans_ver}
    \STATE $ hid\_transf_{ver}: HTV$            \COMMENT{hiding  transfer verifier}         \label{var:token.hid_trans_ver}
    \STATE $ del\_hid\_transf_{ver}: DHTV$      \COMMENT{delegated $^{\prime\prime}$}       \label{var:token.del_hid_trans_ver}
    \STATE $ grabber_{ver}: GV$                 \COMMENT{grabber verifier}                  \label{var:token.grabb_ver}     
\ENDINDENT
\vspace{0.5mm}    
\STATE{ $\textit{\textbackslash*}\; direct\;operations\; \textit{*\textbackslash}$} 
\STATE \textbf{mint}$(t: MT)$            
\STATE \textbf{transfer}$(t: TT)$
\STATE \textbf{revealingTransfer}$(t: RTT)$         
\STATE \textbf{hidingTransfer}$(t: HTT)$               
\vspace{0.5mm}    
\STATE{ $\textit{\textbackslash*}\; delegated\;operations\; \textit{*\textbackslash}$} 
\STATE \textbf{delegatedMint}$(t: DMT)$
\STATE \textbf{delegatedTransfer}$(t: DTT)$
\STATE \textbf{delegatedRevealingTransfer}$(t: DRTT)$
\STATE \textbf{delegatedHidingTransfer}$(t: DHTT)$
\vspace{0.5mm}    
\STATE \textbf{grab}$(t: GT)$
}

For managing balances and NFTs, $balances$ and $nfts$ mappings are used.  
Besides that, the contract relies on several verifier contracts for different tasks.  
$mint\_v$ handles issuance verification, 
$transfer\_v$ manages transfers, 
$revealing\_v$ and $hidden\_v$ deal with revealing and hiding transaction details, 
and $del\_mint\_v$, $del\_transfer\_v$, $del\_rev\_transf\_v$, and $del\_hid\_transf\_v$ are responsible for delegated operations.  
$grabber\_v$ verifies grabber actions.

\subsubsection{Issuance Flow}

Token issuance is the process that creates the digital representation (token) of a real asset on the blockchain platform. 
The life cycle of a token begins with its issuance on the platform and ends with its withdrawal through a burning operation.
Token issuance will be carried out by the issuing authority or by authorized contracts.

\begin{figure}[!ht]
    \centering
    \includegraphics[width=1\linewidth]{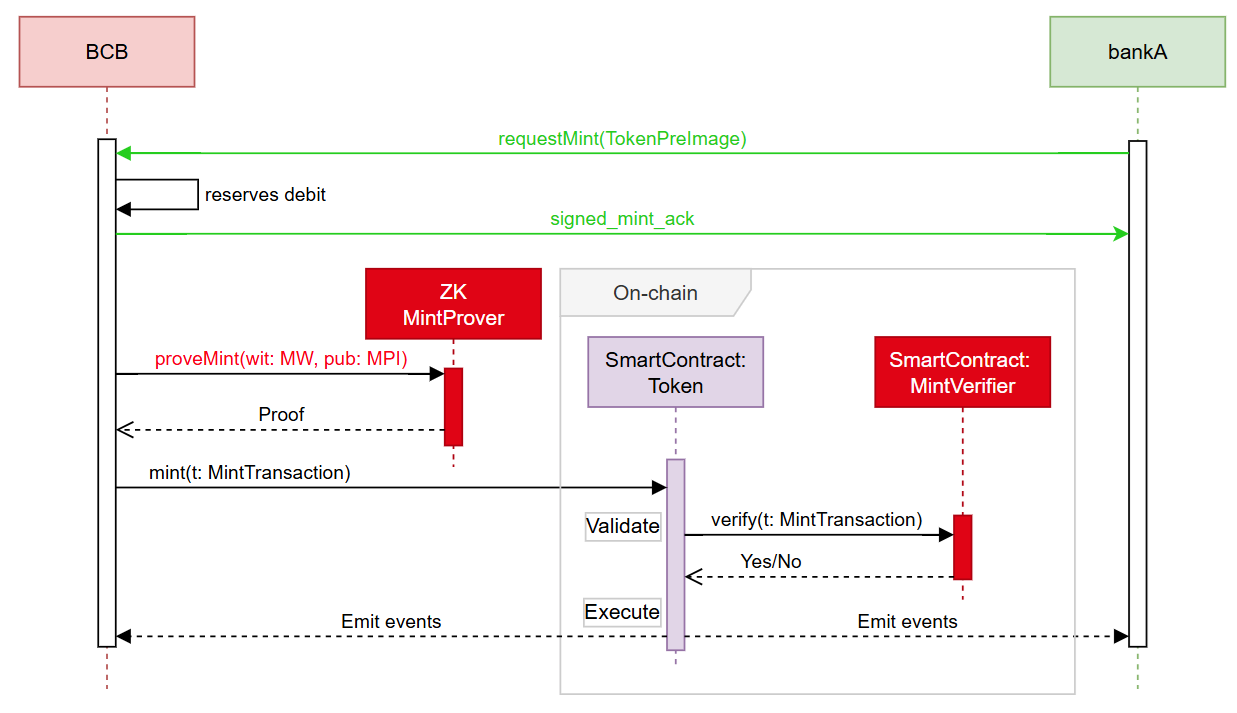}
    \caption{Issuance Flow}
    \label{fig:mint-flow}
\end{figure}

Figure~\ref{fig:mint-flow} presents a sequence diagram that exemplifies the steps of the Token Issuance process.
The process starts with bank A ($bankA$) sending a token issuance request to the Central Bank ($BCB$), i.e., the authority allowed to execute this transaction. 
The BCB realizes the respective debit in bank A's reserve account and responds with a signed confirmation to bank A.
In parallel, the BCB generates a proof ($Proof$) of the issuance based on the witness $MW$ ($MintWitness$) and public inputs $MPI$ ($MintPublicInputs$). 

Using the received proof, the BCB assembles an issuance transaction ($MintTransaction$), composed by the  $MintPublicInputs$ and the $Proof$, then calls the mint function on the Token smart contract. 
The Token contract validates the transaction and requests the issuance verifier smart contract ($MintVerifier$) to verify the validity of the ZK proof contained in the transaction.
If the validations are successful, the transaction is executed, the token is issued, and events related to the token issuance are generated.
If any validation fails, the entire process is reverted, and the transaction is rejected.

Algorithm~\ref{cir:mint} details the zero-knowledge proof (ZK) circuit for minting new tokens. 
This circuit ensures the validity of token issuance while preserving the privacy of sensitive information. 
It utilizes two data structures for witness and public inputs, containing respectively private and public data, to generate a proof that can be verified without revealing the underlying details.

The $MintWitness - MW$ data type defines the structure of the witness. 
It includes a list of $TokenPreImage - TPre$ elements representing the tokens to be minted. 
Optionally, it can also include the issuer's secret key ($issuer_{sk}$) and a Merkle path ($path_i$) for proving the issuer's authorization within a designated issuer tree. 
The corresponding $MintPublicInputs - MPI$ data type comprises the public inputs for the minting process. 
These include the token type ($type_t$), a list of commitments $comms$ corresponding to the tokens being minted, and, optionally, the root of the issuer tree ($root_i$).

The core of the algorithm lies in the $proveMint$ function at line~\ref{cir.mint}. 
This function takes a Mint Witness ($MW$) and a Mint Public Inputs ($MPI$) structures as inputs and returns a ZK proof. 
It begins by enforcing several requirements to ensure the validity of the inputs. 
It checks that the token type in the public inputs matches the type of the tokens being minted, that the number of commitments corresponds to the number of tokens, and that there are no duplicate tokens or commitments. 

The function then iterates through each token in the witness and verifies that its commitment matches the corresponding commitment in the public inputs. 
This step ensures the integrity of the tokens being minted. 
If the optional issuer information is provided, the algorithm calculates the issuer's account from the secret key and verifies its presence in the issuer tree using the provided Merkle path and root.

Finally, the function converts the witness into a ZK proof using the $convertToProof$ function and returns this proof. 
This proof can be used to verify the validity of the token minting process without revealing any of the private information contained in the witness.

\algoritmo{Mint's ZK Circuit}{\label{cir:mint}}{
\INDENT{\textbf{Data Types:}}  
    \STATE $ MW:$                                                   \COMMENT{mint's private witness}                    \label{data:mint_witness}
        \STATE \tab $ outputs: List\langle TPre \rangle,$           \COMMENT{tokens to mint}                            \label{data:mint_wit.token}
        \STATE \tab $ [issuer_{sk}: SKey,]$                         \COMMENT{optional, issuer's secret key}             \label{data:mint_wit.issuer_sk}
        \STATE \tab $ [path_i: MPat]$                               \COMMENT{optional, path in the issuer tree}         \label{data:mint_wit.path}
    \STATE $ MPI:$                                                  \COMMENT{mint's public inputs}                      \label{data:mint_pub_input}
        \STATE \tab $ type_t: uint256,$                             \COMMENT{token type}                                \label{data:mint_pub.type}
        \STATE \tab $ comms: List\langle TCom \rangle,$             \COMMENT{commitments}                               \label{data:mint_pub.commitment}
        \STATE \tab $ [root_i: MRoot]$                              \COMMENT{optional, root of the issuer tree}         \label{data:mint_pub.root_i}
\ENDINDENT
\vspace{2.5mm}
\INDENT{\textbf{proveMint}$(wit: MW, pub: MPI) : uint256$}\label{cir.mint}
    \require $ wit.outputs.size \eq pub.comms.size$                       
    \require $ \forall t \in wit.outputs, t.type \eq pub.type_t $   \COMMENT{correct token type}                        \label{cir.mint.req.type}
    \require $ \nexists\; a,b \in wit.outputs: a \eq b$             \COMMENT{can't repeat tokens}
    \require $ \nexists\; a,b \in pub.comms: a \eq b$               \COMMENT{can't repeat commitments}
    \FOR[check commitments]{ $ i \get 0 \;to\; wit.outputs.size$}
        \STATE   $ c \get commit(wit.outputs[i])$                                                                       \label{cir.mint.req.commit}
        \require $ pub.comms[i] \eq c$                
    \ENDFOR
    \IF[is there a hidden issuer?]{$pub.root_i \ne 0$}
        \require $ wit.issuer_{sk} \ne 0$                           \COMMENT{issuer's secret key is needed}
        \require $ wit.path_i \ne 0$                                \COMMENT{issuer's path is needed}
        \STATE   $ acc \get getAccount(wit.issuer_{sk})$
        \STATE   $ hash \get hash256(acc)$
        \STATE   $ root \get getRoot(hash, wit.path_i)$             \COMMENT{calculate issuer tree root}
        \require $ pub.root_i \eq root$                                                                                \label{cir.mint.req.root}
    \ENDIF
    \RETURN  $ convertToProof(wit)$
\ENDINDENT
}

The \textbf{Mint Transaction} (Algorithm~\ref{alg:mint}) governs the issuance of new tokens within the blockchain platform. 
It leverages zero-knowledge proofs (ZKPs) to ensure the validity and integrity of the minting process while maintaining the confidentiality of sensitive information. 
This contract defines the rules and procedures for creating new tokens and manages the associated data structures.

The contract first defines the $MintTransaction$ data type (line~\ref{data:mint_transaction}), which encapsulates the information required for a minting operation. 
This includes the {$MintPublicInputs$} (Algorithm~\ref{cir:mint}, line~\ref{data:mint_pub_input}), containing the public parameters of the minting process, and a $proof$, which is a ZKP that validates the transaction. 

The core functionality of the contract is provided by the $mint$ function at line~\ref{alg.mint}. 
This function allows  authorized entities to mint new tokens. 
It requires that the transaction be initiated by either a registered public issuer or a private issuer with a valid proof of its presence in the issuer tree. 
The function then invokes the $mint_{v}.verify$ function to verify the ZKP provided in the transaction, ensuring its validity. 
If the verification is successful, the $doMint$ function is called to execute the minting operation.

The $doMint$ function (line~\ref{alg.domint}) performs the actual token creation and updates the contract's state accordingly. 
Before, it checks that the token type in the transaction matches the contract's designated token type and that the commitments associated with the new tokens are not already present in the commitment tree. 
If the transaction involves a private issuer, it verifies that the issuer is authorized by checking her presence in the private issuer tree. 
Then, the function adds the new token commitments to the commitments tree. 
Finally, the function $mint$ emits events to signal the successful completion of the minting process.       

\algoritmo{- TK Mint Transaction}{\label{alg:mint}}{
\INDENT{\textbf{Data Types:}}  
    \STATE $MT:$                        \COMMENT{Mint transaction}                                      \label{data:mint_transaction}
        \STATE \tab $pub: MPI,$         \COMMENT{public Mint inputs -- alg.~\ref{cir:mint}}             \label{data:mint_tras.pub}
        \STATE \tab $proof: uint256$    \COMMENT{Mint proof}                                            \label{data:mint_tras.proof}
\ENDINDENT

\vspace{2.5mm}
\INDENT[mint tokens]{\textbf{mint}$(t: MT)$}\label{alg.mint}  
    \STATE   $ public_i \get this.issuers[msg.sender]$
    \STATE   $ private_i \get t.pub.root_i \ne \bot$    
    \require $ public_i \xor private_i$                 \COMMENT{or public or private issuer}           \label{alg.mint.req.issuer}
    \require $ this.mint_v.verify(t)$                   \COMMENT{verify ZKP}                            \label{alg.mint.req.proof}
    \STATE   $ doMint(t.pub)$                                                                           \label{alg.mint.do}  
    \events
\ENDINDENT
\vspace{2.5mm}
\INDENT{\textbf{doMint}$(pub: MPI)$} \label{alg.domint}  
    \require $ pub.type_t \eq this.type_t$                          \COMMENT{token type OK}             \label{alg.domint.req.type}
    \require $ \forall c \in pub.comms,  c \not\in this.tree_c$     \COMMENT{only new commitments}      \label{alg.domint.req.commitment}  
    \IF [from a hidden issuer?]{$ t.root_i \ne \bot$}
        \require $ pub.root_i \in this.tree_i.roots$                \COMMENT{issuer's tree root OK}     \label{alg.domint.req.r.root_c}
    \ENDIF
    \STATE $ this.tree_c \cupeq t.pub.comms$                        \COMMENT{store commitments}         \label{alg.domint.add.commitment}  
\ENDINDENT
}

\subsubsection{Transfer/Burn Transaction}\label{sec:transfer_burn}

Once a token has been issued and its commitment (a cryptographic representation of the token, section~\ref{sec:token_commitment}) registered in the smart contract's commitments Merkle tree, the most basic operations participants can perform are transfer and withdrawal.
Transfer operations change the ownership and possession of tokens within the blockchain platform, without any interference or changes to the participants' reserve accounts.
Participants can also transfer their tokens to smart contracts, but this type of transfer will be discussed in a later section of this document.
Just as the full issuance flow debits the value from the bank's reserve account and mints a token of that value on the blockchain, the full withdrawal flow burns a specific amount of tokens on the blockchain and, upon notification of this token destruction, the authority credits the corresponding value to the reserve account.

The option to unify the transfer and burn operations into a single blockchain transaction allows for partial withdrawal of consumed tokens, if desired. 
An additional benefit of this unified transaction is that the presence of the burn commitment, even when the withdrawal amount is zero, prevents distinguishing between simple transfers and those associated with a withdrawal, enhancing blockchain privacy even more.

\begin{figure}[!ht]
    \centering
    \includegraphics[width=1\linewidth]{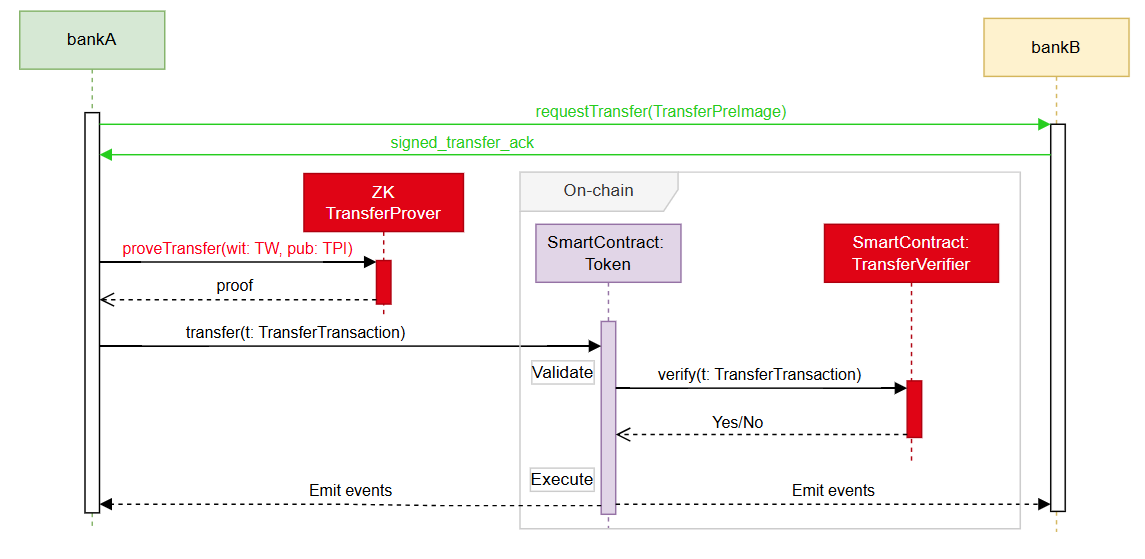}
    \caption{Transfer Flow}
    \label{fig:transfer.flow}
\end{figure}

\begin{figure}[!ht]
    \centering
    \includegraphics[width=1\linewidth]{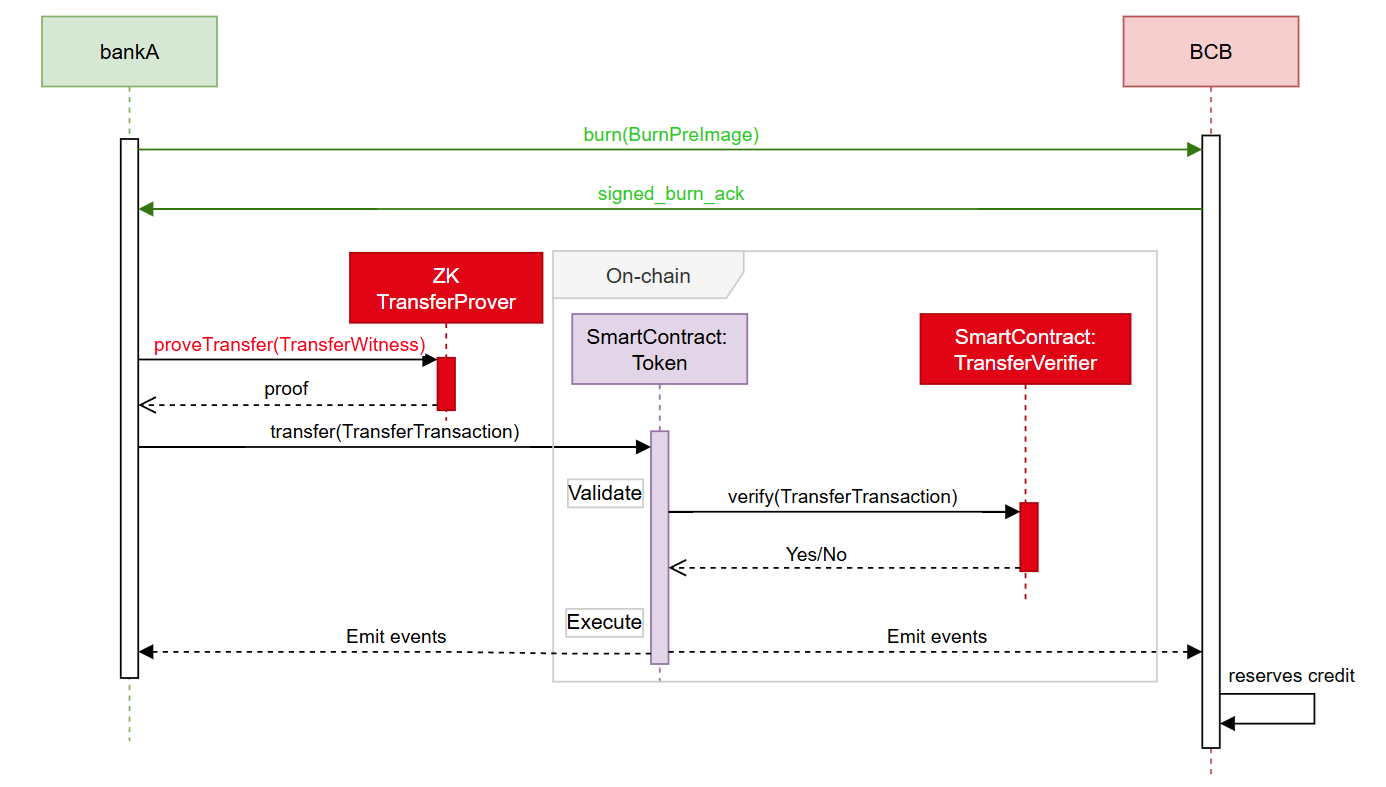}
    \caption{Withdrawal Flow}
    \label{fig:burn.flow}
\end{figure}

Figures~\ref{fig:transfer.flow} and~\ref{fig:burn.flow} illustrate the transfer and withdrawal flows, respectively. 
These flows can happen simultaneously, using the same blockchain transaction, the only requirement for that is to execute the two private communications.
In both scenarios, Bank A initiates the process by sending a request through a private channel to the receiving party. 
This request includes either a $TransferPreImage$ for transfers (Figure~\ref{fig:transfer.flow} and Algorithm~\ref{alg:trans_data}) or a $BurnPreImage$ for withdrawals (Figure~\ref{fig:burn.flow} and Algorithm~\ref{alg:burn_data}). 
The receiving party, which is either Bank B for transfers or the Central Bank for withdrawals, performs internal checks and validations. 
If approved, the receiving party sends a signed acknowledgment to Bank A, which must store this approval for audit purposes. 

Bank A then uses a zero-knowledge transfer prover circuit to generate a proof based on the required parameters, including the transfer's/burn's witness ($TW$) and public inputs ($TPI$). 
This proof, along with its public inputs, is submitted to the token smart contract ($TK$) as a transfer transaction ($TT$). 
The token contract validates the transaction while a separate transfer verifier smart contract independently verifies the proof. 
Upon successful verification, the token contract executes the requested action – transferring tokens to Bank B in the transfer flow, and/or burning tokens in the withdrawal flow. 
In both cases, the contract emits events to reflect the updated state. 
Finally, in the withdrawal flow (Figure~\ref{fig:burn.flow}), the corresponding reserves are credited to Bank A.

The transfer's preimage ($TransferPreImage$) is defined at Algorithm~\ref{alg:trans_data}.
It contains the information required by the receiver to authorize the transaction and consume its outputs when needed. 
It has two list fields in its structure: 
$outputs$, the complete list of receiver's tokens' preimages created by this transfer, only the receiver's tokens will be shown here, and an optional field, $inputs$, with the 
consumed token's nullifiers' preimage ($NPre$).
This optional field is required when the receiver institution needs to verify the payload information of the consumed tokens.
The Nullifier preimage has a peculiar structure composed of two Fields, the $input\_payload$, a clear copy of the respective consumed token preimage payload, and the $partial\_hash$, which is the hash of all the other fields of the nullified token preimage (Algorithm~\ref{cir:token_commitment}).
 The payload is the unique field from the consumed tokens that will be revealed to the receiver institution. 
 This way, the receiver participant is able to read the payload information of the consumed tokens and execute its internal checks, then rebuild the nullifier commitment in order to recognize this transaction when it is processed by the blockchain.

\data{Transfer preimage}{\label{alg:trans_data}}{
\STATE $TransferPreImage:$                              \COMMENT{transfer's preimage}              \label{data:trans_pre}
    \STATE \tab $outputs: List\langle TPre \rangle,$    \COMMENT{receiver's token preimage}        \label{data:trans_pre.output}
    \STATE \tab $[inputs: List\langle NPre \rangle]$    \COMMENT{optional, nullifiers preimage}    \label{data:trans_pre.nullifier}
\STATE $NPre:$                                          \COMMENT{Nullifier preimage}               \label{data:null_pre}    
    \STATE \tab $partial\_hash: uint256,$               \COMMENT{nullifier's partial hash}          \label{data:null_pre.hash}
    \STATE \tab $input\_payload: any$                   \COMMENT{input payload}                     \label{data:null_pre.payload}
}

The $BurnPreImage$ structure in Algorithm~\ref{alg:burn_data} encapsulates the information required to request tokens withdrawal to the token authority. 
It encapsulates the necessary information to allow the token authority to confirm the burn commitment inside the transfer/burn transaction in the blockchain.
It comprises three fields:  
$amount$ specifies the amount of fungible tokens to be burned, represented as a 128-bit unsigned integer. 
$ids$ that lists the identifiers of the NFTs to be burned. 
Finally, $nonce$ which is a 256-bit unsigned integer serving as the source of randomness, ensuring each burn operation is distinct.
The $amount$ and $ids$ fields are optional, which means that they can be set to zero or $\varnothing$ if not used.

\data{Burn preimage}{\label{alg:burn_data}}{
\STATE $BurnPreImage:$                                      \COMMENT{Burn preimage}                    \label{alg:burn_pre}
    \STATE \tab $[amount: uint256,]$                        \COMMENT{fungible amount to burn}    \label{alg:burn_pre.amount}
    \STATE \tab $[ids: List\langle uint256 \rangle,]$       \COMMENT{NFT ids to burn}            \label{alg:burn_pre.ids}    
    \STATE \tab $nonce: Nonce$                              \COMMENT{Burn entropy}                      \label{alg:burn_pre.nonce}
}

Algorithms~\ref{cir:trans} to~\ref{cir:trans_aux_2} describe the zero-knowledge circuit for the transfer/burn functionality. 
This circuit ensures the secure and private transfer of tokens between users while also allowing for the burn of tokens with the same properties. 
The ZK validation is executed off-ledger and guarantees the knowledge of private fields and their relation with themselves and the public inputs.

The transfer witness ($TW$) data type defines the structure of the private data used in the proof. 
It includes a list of Image and Path pairs ($Img\_Path$) named $inputs$, each containing a token preimage ($TPre$) $img$ and a Merkle path ($MPat$) $path$ to prove its presence in the commitment tree. 
It also includes the $outputs$, a list of token preimages ($TPre$), the sender's secret key ($SKey$) $sk$, the auditor's public key ($PKey$) $audit_{pk}$ and two optional fields\footnote{Optional fields can be settled to zero or $\varnothing$ if not used.}, the amount to be burned $burn_a$ and the NFT identifiers to burn $burn_{ids}$. 

The transfer public inputs ($TPI$) data type contains the list of input tokens' nullifiers ($TNul$), named $nulls$, the list of input tokens' grabbers ($TGra$), named $grabs$ and a list of created tokens' commitments ($TCom$), named $comms$. 
The token smart contract's type $token_t$, its Merkle tree root ($MRoot$) $root_c$, grabber nonce $nonce_g$ and auditor's account $audit_{acc}$. 
Beyond that, the burn commitment\footnote{Even if there is nothing to burn, the burning commitment will be present.} $burn_c$ and audit data $audit_d$. 

The $proveTransfer$ function, which receives a transfer witness $wit$ and public inputs $pub$, orchestrates the verification process. 
It calls several auxiliary functions to check the validity of the inputs and outputs, ensure mass conservation, validate the burn process, and verify the audit data.
These auxiliary functions are defined in Algorithms~\ref{cir:trans_aux} and~\ref{cir:trans_aux_2}.
Once all checks are successfully performed, the $proveTransfer$ function creates and returns the ZK proof that will be used by the verifier instead of the witness; this process is represented by the hypothetical function $convertToProof$. 

\circuit{Transfer/Burn}{\label{cir:trans}}{
\INDENT{\textbf{Data Types:}}    
    \STATE $Img\_Path : \{img: TPre, path: MPat\}$ \label{data:img_and_path}
    \STATE $TW:$                                                                    \COMMENT{transfer witness}          \label{data:trans_witness}  
        \STATE \tab $inputs: List\langle Img\_Path \rangle,$                                                            \label{data:trans_wit_inputs}
        \STATE \tab $outputs: List\langle TPre\rangle,$                                                                 \label{data:trans_wit_outputs} 
        \STATE \tab $sk: SKey,$                                                     \COMMENT{payer's secret key}        \label{data:trans_wit_sk}
        \STATE \tab $audit_{pk}: PKey,$                                             \COMMENT{audit's public key}        \label{data:trans_wit_audit_pk}
        \STATE \tab $[burn_a: uint256,]$                                            \COMMENT{amount to burn}            \label{data:trans_wit_burn_amount}
        \STATE \tab $[burn_{ids}: List\langle uint256 \rangle]$                     \COMMENT{ids to burn}               \label{data:trans_wit_burn_ids}
    \STATE $TPI:$                                                                   \COMMENT{transfer's public inputs}  \label{data:trans_public_inputs}
        \STATE \tab $nulls: List\langle TNul \rangle,$                              \COMMENT{nullifiers}                \label{data:trans_pub_nullifiers}
        \STATE \tab $grabs: List\langle TGra \rangle,$                              \COMMENT{grabbers}                  \label{data:trans_pub_grabbers}
        \STATE \tab $comms: List\langle TCom \rangle,$                              \COMMENT{commitments}               \label{data:trans_pub_commitments}  
        \STATE \tab $type_t: uint256,$                                              \COMMENT{token type}                \label{data:trans_pub_type}  
        \STATE \tab $root_c: MRoot,$                                                \COMMENT{commitments' tree root}    \label{data:trans_pub_root_c}
        \STATE \tab $nonce_g: Nonce,$                                               \COMMENT{TK's grabber nonce}        \label{data:trans_pub_grab_nonce}
        \STATE \tab $burn_c: uint256,$                                              \COMMENT{burn commitment}           \label{data:trans_pub_burn_commit}
        \STATE \tab $audit_{acc}: Account$                                          \COMMENT{auditor's account}         \label{data:trans_pub_auth_acc}
        \STATE \tab $audit_d: Bytes,$                                               \COMMENT{audit data, algorithm~\ref{cir:token_nullifier}}\label{data:trans_pub_audit_data}
\ENDINDENT
\vspace{2.5mm}
\INDENT{\textbf{proveTransfer}$(wit: TW, pub: TPI) : uint256$}     
    \require $checkInputs(wit,pub)$                                 \COMMENT{correct inputs}        
    \require $checkOutputs(wit, pub)$                               
    \require $checkMassConservation(wit)$                           \COMMENT{conserved mass}         
    \require $checkBurn(wit, pub)$                                  \COMMENT{correct withdrawal}     
    \require $checkAuditData(wit, pub)$                             \COMMENT{audit data}            
    \RETURN  $convertoToProof(wit)$                                 \COMMENT{returns ZK proof}
\ENDINDENT
}

The code at Algorithm~\ref{cir:trans_aux} defines two crucial functions, $checkInputs$ and $checkOutputs$, within the zero-knowledge circuit for Transfer/Burn. 
These functions are responsible for validating the inputs and outputs of a transfer/burn transaction within the ZK circuit.

The $checkInputs$ function, defined at line~\ref{cir.check_inputs}, verifies the validity of the inputs used in the transaction. 
It takes a witness $wit$ of type $TW$ and public inputs $pub$ of type $TPI$ as parameters. 

It first checks if the number of nullifiers $pub.nulls$ and grabbers $pub.grabs$ matches the number of inputs provided in the witness. 
Also ensures that there are no duplicate input preimages within the witness inputs set (line~\ref{cir.check_inputs_req_img}). 
It then derives the payer's grabber key $grab_k$ using the function $createGrabberKey$ (Algorithm~\ref{cir:token_grabber}), the payer's secret key ($wit.sk$) and the public TK's grabber nonce ($pub.nonce_g$). 

The function then iterates through each witness input in $inputs$. 
For each input, it retrieves the image ($img_{in}$) and the path ($path_{in}$) from the $inputs$ array.
It computes the commitment ($comm_{in}$) by applying the $commit$ function (Algorithm~\ref{cir:token_commitment}) to the input token preimage. 
It also calculates the nullifier ($null_{in}$) and grabber ($grab_{in}$) using the $nullify$ (Algorithm~\ref{cir:token_nullifier}) and $grab$ (Algorithm~\ref{cir:token_grabber}) functions, respectively. 
The $nullify$ function uses the preimage and payer's secret key, while the $grab$ function uses the preimage and the payer's grabber key ($grab_k$).    
Then It calculates the root of the commitments Merkle tree ($root_{in}$) using the $getRoot$ function (Algorithm~\ref{cir:merkle_root}) with the newly calculated commitment ($comm_{in}$) and the extracted token's path ($path_{in}$) (line~\ref{cir.check_inputs_root}). 

Finally, it asserts several conditions:         
the input image must have either a non-zero amount or a non-zero ID (line~\ref{cir.check_inputs_req_zero}); 
the input image type must match the public transaction token type (line~\ref{cir.check_inputs_req_type}); 
the computed nullifier must match the corresponding public nullifier in $pub.nulls$ (line~\ref{cir.check_inputs_req_null}); 
the computed grabber must match the corresponding public grabber in $pub.grabs$ (line~\ref{cir.check_inputs_req_grab}); 
and the computed root must match the public root $pub.root_c$ (line~\ref{cir.check_inputs_req_root}). 
If all conditions are met for all inputs, the function returns true.

The $checkOutputs$ function, defined at line~\ref{cir.check_outputs}, validates the outputs generated by the transaction. 
It accepts a transfer witness ($wit$) and a transfer public inputs ($pub$) as input. 
Initially, it verifies that the number of output preimages in $wit.outputs$ matches the number of commitments in $pub.comms$. 
It also checks for duplicates within both the $wit.outputs$ array and the $pub.comms$ array.
The function then iterates through each output in $wit.outputs$. 

For each output, it retrieves the preimage ($img_{out}$) from the $wit.outputs$ array and computes its commitment ($comm_{out}$) using the $commit$ function (Algorithm~\ref{cir:token_commitment}). 
It determines if the output represents a fungible token ($isFT$) by checking if $img_{out}.amount$ is non-zero and similarly determines if it represents a nonfungible token ($isNFT$) by checking if $img_{out}.id$ is non-zero. 
It then requires that each output be either a fungible or a nonfungible token. 
Additionally, it checks that the output image type matches the public transaction type $pub.type_t$ and that the computed commitment matches the corresponding public commitment in $pub.comms$. 
If all conditions are satisfied for all outputs, the function returns true.

\circuit{Transfer/Burn - \textit{continuation}}{\label{cir:trans_aux}}{
\INDENT{\textbf{checkInputs}$(wit: TW, pub: TPI): bool$}    \label{cir.check_inputs}
    \require $ wit.inputs.size \eq pub.nulls.size$                         
    \require $ wit.inputs.size \eq pub.grabs.size$                           
    \require $ \nexists\; a,b \in \{i.img, \forall i \in wit.inputs\}: a \eq b$\label{cir.check_inputs_req_img}
    \STATE $ grab_k \get createGrabberKey(wit.sk, pub.nonce_g)$         
    \FOR[for each input]{$i\get 0 \;to\; wit.inputs.size$}                                           
        \STATE   $ img_{in} \get wit.inputs[i].img$             \label{cir.check_inputs_img}     \COMMENT{get image}
        \STATE   $ path_{in} \get wit.inputs[i].path$           \label{cir.check_inputs_path}    \COMMENT{get path}
        \STATE   $ comm_{in} \get commit(img_{in})$             \label{cir.check_inputs_comm}    \COMMENT{calculate commitment}
        \STATE   $ null_{in} \get nullify(img_{in}, wit.sk)$    \label{cir.check_inputs_null}    \COMMENT{calculate nullifier}    
        \STATE   $ grab_{in} \get grab(img_{in}, grab_k)$       \label{cir.check_inputs_grab}    \COMMENT{calculate grabber}
        \STATE   $ root_{in} \get getRoot(comm_{in}, path_{in})$\label{cir.check_inputs_root}    \COMMENT{calculate tree root}
        \require $ img_{in}.amount \ne 0 \lor img_{in}.id \ne 0$\label{cir.check_inputs_req_zero}
        \require $ img_{in}.type \eq pub.type_t$                \label{cir.check_inputs_req_type}                
        \require $ null_{in} \eq pub.nulls[i]$                  \label{cir.check_inputs_req_null}           
        \require $ grab_{in} \eq pub.grabs[i]$                  \label{cir.check_inputs_req_grab}             
        \require $ root_{in} \eq pub.root_c$                    \label{cir.check_inputs_req_root}                
    \ENDFOR
    \RETURN \TRUE
\ENDINDENT
\vspace{2.5mm}
\INDENT{\textbf{checkOutputs}$(wit: TW, pub: TPI): bool$}   \label{cir.check_outputs}
    \require $ wit.outputs.size \eq pub.comms.size$           
    \require $ \nexists\; a,b \in wit.outputs: a \eq b$         \COMMENT{no duplicated output}
    \require $ \nexists\; a,b \in pub.comms: a \eq b$           \COMMENT{no duplicated commitment}
    \FOR[for each output]{$i\get 0 \;to\; wit.outputs.size$}                               
        \STATE   $ img_{out} \get wit.outputs[i]$               \label{cir.check_outputs_}       \COMMENT{get preimage}
        \STATE   $ comm_{out} \get commit(img_{out})$           \label{cir.check_outputs_1}      \COMMENT{calculate commitment}    
        \STATE   $ isFT \get img_{out}.amount \ne 0$            \label{cir.check_outputs_2}      \COMMENT{is fungible}
        \STATE   $ isNFT \get img_{out}.id \ne 0$               \label{cir.check_outputs_3}      \COMMENT{is nonfungible}
        \require $ isFT \lor isNFT$                             \label{cir.check_outputs_4}      \COMMENT{or fungible or nonfungible}
        \require $ img_{out}.type \eq pub.type_t$               \label{cir.check_outputs_5}      \COMMENT{type OK}
        \require $ comm_{out} \eq pub.comms[i]$                 \label{cir.check_outputs_6}      \COMMENT{commitment OK}
    \ENDFOR
    \RETURN \TRUE
\ENDINDENT
}

Algorithm~\ref{cir:trans_aux_2} presents three additional functions, $checkBurn$, $checkMassConservation$, and $checkAuditData$, which are part of the zero-knowledge circuit for transfer/burn operations in the $TK$ smart contract. 
These functions collectively enforce important constraints within the Transfer/Burn circuit, ensuring that burn operations are valid, mass is conserved, and audit data is correctly generated and verified. 

The $checkBurn$ function, defined at line~\ref{cir.check_burn}, verifies the validity of the burn operation within a transaction. 
It accepts a witness $wit$ of type $TW$ and public inputs $pub$ of type $TPI$ and returns a boolean indicating success or failure. 
First, it checks that no NFT ID is repeated within the $wit.burn_{ids}$ set. 
Then, it initializes an empty set $burn_{items}$. 
If $wit.burn_a$ is not zero, it is added to $burn_{items}$. 
If $wit.burn_{ids}$ is not empty, its contents are added to $burn_{items}$. 
After that, it calculates $wit_{hash}$ as the 256 bit hash of the entire witness $wit$. 
Finally, it computes $burn\_c$ as the 256 bit hash of $burn_{items}$ and $wit_{hash}$. 
The function returns true if $pub.burn_c$ is equal to the calculated $burn\_c$, indicating a correct burn operation.

The $checkMassConservation$ function (lines~\ref{cir.check_mass_conservation}--\ref{cir.check_mass_conservation_return}) verifies that the total amount and IDs of assets are conserved by the transaction. 
It takes a transfer witness $wit$ as input. 
The function calculates the total amount of fungible tokens being consumed ($total_{in}$) by summing the amounts in the preimage of $inputs$ and similarly calculates the total amount of fungible tokens being created ($total_{out}$) by summing the amounts in the $outputs$. 
It then verifies that the total input amount equals the total output amount plus the burned fungible amount $burn_a$, ensuring conservation of fungible tokens. 
It further extracts the sets of nonfungible token IDs being consumed ($ids_{in}$) and created ($ids_{out}$) from the inputs and outputs respectively.
It then checks that the set of consumed nonfungible token IDs is equal to the union of the set of created IDs and the set of burned IDs $burn_{ids}$, enforcing conservation of nonfungible tokens. 
Finally, if all checks are OK, the function returns true.

The $checkAuditData$ function, defined at line~\ref{cir.check_audit_data}, is responsible for generating and verifying audit data associated with the transaction. 
It accepts a transfer witness $wit$ and a transfer public inputs $pub$ as input. 
It prepares the data to be audited ($audit\_pre\_img$) by conditionally including input images, output images, burned fungible amount, and burned nonfungible IDs, only if they are present in the witness. 
Then, it verifies that the public input field $audit_{acc}$ is the 256 bit hash of the witness's audit public key $audit_{pk}$. 
Finally, it requires that the public audit data $audit_d$ matches the encrypted value of $audit\_pre\_img$ using the audit public key $audit_{pk}$ as the encryption key and returns true if all conditions hold.

\circuit{Transfer/Burn - \textit{continuation}}{\label{cir:trans_aux_2}}{
\INDENT{\textbf{checkBurn}$(wit: TW, pub: TPI): bool$}  \label{cir.check_burn}
    \require $ \nexists\; a,b \in wit.burn_{ids}: a \eq b$                         \COMMENT{can't repeat id}
    \STATE $burn_{items} \get \varnothing $
    \IF [burning FT?]{$ wit.burn_a \ne 0 $}
        \STATE $burn_{items} \peq wit.burn_a$                                     \COMMENT{add amount to burn items}
    \ENDIF   
    \IF [burning NFT?]{$ wit.burn_{ids} \ne \varnothing$}
        \STATE $burn_{items} \cupeq wit.burn_{ids}$                               \COMMENT{add ids to burn items}
    \ENDIF
    \STATE  $ wit_{hash} \get hash256(wit)$
    \STATE  $ burn\_c \get hash256(burn_{items}, wit_{hash})$
    \RETURN $ pub.burn_c \eq burn\_c$                                              \COMMENT{correct burn}   \label{cir.check_burn_return}
\ENDINDENT
\vspace{2.5mm}
\INDENT{\textbf{checkMassConservation}$(wit: TW): bool$}  \label{cir.check_mass_conservation}     
    \STATE   $ imgs_{in} \get \{in.img,\;\forall in \in wit.inputs\}$
    \STATE   $ total_{in} \get getAmountSum(imgs_{in})$                         \COMMENT{fungible consumed}
    \STATE   $ ids_{in} \get \{i.ids,\;\forall i \in imgs_{in}\}$               \COMMENT{nonfungible consumed}    
    \STATE   $ imgs_{out} \get \{out.img,\;\forall out \in wit.outputs\}$        
    \STATE   $ total_{out} \get getAmountSum(imgs_{out})$                       \COMMENT{fungible created}
    \STATE   $ ids_{out} \get \{i.ids,\;\forall i \in imgs_{out}\}$             \COMMENT{nonfungible created}    
    \require $ total_{in} \eq total_{out} + wit.burn_a$                         \COMMENT{mass conserved}
    \require $ ids_{in} \eq ids_{out} \cup wit.burn_{ids}$                      \COMMENT{mass conserved}
    \RETURN \TRUE       \label{cir.check_mass_conservation_return}
\ENDINDENT
\vspace{2.5mm}
\INDENT{\textbf{getAmountSum}$(imgs : List\langle TPre\rangle): uint256$} \label{cir.get_amount_sum}
    \RETURN $ sum(\{i.amount,\;\forall i \in imgs\})$  
\ENDINDENT
\vspace{2.5mm}
\INDENT{\textbf{checkAuditData}$(wit: TW, pub: TPI): bool$} \label{cir.check_audit_data}
    \STATE   $ imgs_{in} \get wit.inputs \ne \bot ?\; \{i.img, \forall i \in wit.inputs \} : \varnothing$ 
    \STATE   $ imgs_{out} \get wit.outputs \ne \bot ?\; wit.outputs : \varnothing$ 
    \STATE   $ burn_{ids} \get wit.burn_{ids} \ne \bot ?\; wit.burn_{ids} : \varnothing$     
    \STATE   $ burn_a \get wit.burn_a \ne \bot ?\; wit.burn_a : 0$
    \STATE   $ audit\_pre\_img \get \{imgs_{in}, imgs_{out}, burn_a, burn_{ids}\}$
    \require $ pub.audit_{acc} \eq hash256(wit.audit_{pk})$
    \require $ pub.audit_d \eq wit.audit_{pk}.cipher(audit\_pre\_img)$       \label{cir.check_audit_data_return}
    \RETURN \TRUE
\ENDINDENT
}

Algorithm~\ref{alg:trans} presents the transfer related functions and data types in $TK$ smart contract, responsible for managing the transfer and withdrawal of tokens within the system. 
These functions work together with the zero-knowledge proofs (ZKPs) circuits related to the transfer/burn flow to ensure the privacy and security of these operations. 
It defines the data structures and functions required to process transfer and withdrawal requests, validate transactions, and update the contract's state accordingly.

The algorithm begins by defining the Transfer Transaction data type ($TT$), which encapsulates the necessary information for a transfer and/or withdrawal operation. 
This includes the Transfer Public Inputs ($TPI$), containing the public parameters of the transaction, and a $proof$, which is a zk-SNARK used for transaction validation. 

The core functionality of the contract is provided by the $transfer$ function. 
This is the public function that processes incoming transfer transactions. 
It first calls the transfer verifier $verify$ function to check the ZKP provided in the transaction, ensuring its validity. 
If the proof is valid, the $doTransfer$ function is invoked to execute the transfer.

The $doTransfer$ function performs the necessary checks and state updates to complete the transfer. 
It ensures that the audit account, token type, and grabber nonce in the transaction match the correspondent contract's values. 
It also verifies that the root of the commitment tree used in the transaction is valid and that there are no duplicate nullifiers, grabbers, or commitments, preventing double-spending and other fraudulent activities. 
Finally, the function updates the contract's state by adding the nullifiers and grabbers to their respective sets and adding the new commitments to the commitment tree.

The $doTransfer$ function performs the necessary checks and state updates to complete the transfer. 
It ensures that the audit account ($pub.audit_{acc}$), token type ($pub.type_t$), and grabber nonce ($pub.nonce\_g$) in the transaction match the corresponding contract's values (lines~\ref{alg.dotrans_req_size}--\ref{alg.dotrans_req_grabb_nonce}). 
It also verifies that the root of the commitment tree used in the transaction ($pub.root\_c$) is valid (line~\ref{alg.dotrans_req_root_c}) and that all the transaction's nullifiers, grabbers, and commitments are new to the smart contract, preventing double-spending and other fraudulent activities (lines~\ref{alg.dotrans_req_nullifiers}--\ref{alg.dotrans_req_outputs}). 
Finally, the function updates the contract's state by adding the nullifiers and grabbers to their respective sets (lines~\ref{alg.dotrans_nullifiers_add} and~\ref{alg.dotrans_grabbers_add}) and adding the new commitments to the commitment tree (line~\ref{alg.dotrans_outputs_add}).

\contract{Transfer/Burn}{\label{alg:trans}}{
\INDENT{\textbf{Data Types:}}   
    \STATE $TT:$                      \COMMENT{transfer's transaction}    \label{data:trans} 
        \STATE \tab $pub: TPI,$       \COMMENT{transfer's public inputs}  \label{data:trans_tranfer}
        \STATE \tab $proof: uint256$  \COMMENT{ZKP}                       \label{data:trans_proof}
\ENDINDENT

\vspace{2.5mm}    
\INDENT{\textbf{+ transfer}$(t: TT)$}     \label{alg.trans}
    \require $this.transfer_v.verify(t)$                   \COMMENT{verify the zk-SNARKS}       \label{alg.trans_req_verify}
    \STATE   $doTransfer(t.transfer)$                                                           \label{alg.trans_do}  
    \events                                                                                     \label{alg.trans_events}  
\ENDINDENT

\vspace{2.5mm}    
\INDENT{\textbf{- doTransfer}$(pub: TPI)$}                                  \label{alg.dotrans}
    \require $ pub.audit_{acc} \eq this.audit\_acc$                         \label{alg.dotrans_req_size}
    \require $ pub.type_t \eq this.type_t$                                  \label{alg.dotrans_req_type}
    \require $ pub.nonce_g \eq this.grab\_nonce$                            \label{alg.dotrans_req_grabb_nonce}
    \require $ pub.root_c \in this.tree_c.roots$                            \label{alg.dotrans_req_root_c}       
    \require $ \forall n \in pub.nulls, n \notin this.nullifiers$           \label{alg.dotrans_req_nullifiers}
    \require $ \forall g \in pub.grabs, g \notin this.grabbers$             \label{alg.dotrans_req_grabbers}
    \require $ \forall o \in pub.outputs, o \notin this.tree_c$             \label{alg.dotrans_req_outputs}
    \STATE   $ \forall n \in pub.nulls, this.nullifiers[n] \get \TRUE$      \label{alg.dotrans_nullifiers_add}
    \STATE   $ \forall g \in pub.grabs, this.grabbers[g] \get \TRUE$        \label{alg.dotrans_grabbers_add}
    \STATE   $ this.tree_c \cupeq  pub.outputs$                             \label{alg.dotrans_outputs_add}
\ENDINDENT
}

\subsubsection{Revealing Transfer Transaction}\label{sec:revealing_transfer}

Revealing transfers are designed to facilitate direct interactions between participants and smart contracts while maintaining a degree of privacy. 
Unlike fully transparent and fully hidden transactions, revealing transfers allow for one or more output tokens to be publicly disclosed on the blockchain, without revealing sensitive information about the consumed input tokens, the sender's identity and the undisclosed outputs. 
This mechanism addresses the current limitation where smart contracts often require full visibility of token data to operate freely, due to their lack of ability to generate or maintain zero-knowledge proofs without compromising privacy.

\begin{figure}[!ht]
    \centering
    \includegraphics[width=1\linewidth]{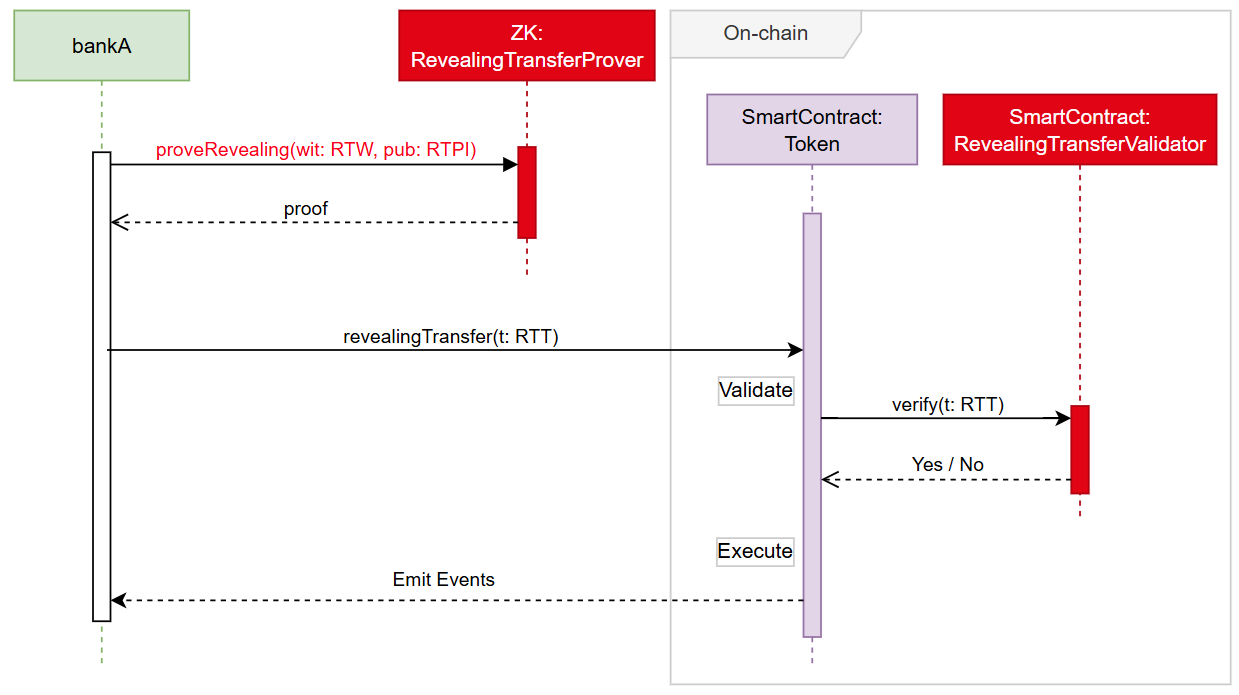}
    \caption{Revealing transfer flow}
    \label{fig:revealing.transfer.flow}
\end{figure}

The sequence diagram at Figure~\ref{fig:revealing.transfer.flow} illustrates the process of a revealing transfer within a blockchain environment. 
The system involves four primary entities: $bankA$, a ZK $RevealingTransferProver$ circuit, an on-chain $Token$ smart contract, and an on-chain $RevealingTransferValidator$ smart contract.

The process initiates with $bankA$ triggering the $RevealingTransferProver$ by invoking $proveRevealing$ with two arguments, of types $RTW$ that represents the "Revealing Transfer Witness", containing private information about the transfer, and $RTPI$ the "Revealing Transfer Public Input" containing the transfer's publicly verifiable information. 
The $RevealingTransferProver$ computes a zero-knowledge proof based on these inputs, demonstrating the validity of the transfer without revealing the sensitive details contained in its witness. 
After receiving the ZK proof, $bankA$ submits a $RevealingTransferTransaction$ the the $revealingTransfer$ function of the $Token$ smart contract. 
This transaction likely encapsulates the public details of the transfer and the generated $proof$.

The $Token$ smart contract, upon receiving the $revealingTransfer$ call, initiates a validation request by calling $verify$ on the $RevealingTransferValidator$ smart contract.
The $RevealingTransferValidator$ smart contract executes the verification logic, assessing the validity of the transaction based on the $proof$ and the transaction's public details.
If the validator contract accepts the ZK proof, the Token smart contract executes the transfer, updating the on-chain token balances accordingly, otherwise the transaction is rejected and all changes are reverted. 
Finally, the $Token$ contract emits relevant events, signaling the completion of the transfer.

\noindent \textit{Note: To enhance readability, this section omits the fields, data types, and functions related to token burning. These features can be implemented by adapting the mechanisms described in Section~\ref{sec:transfer_burn}.}

Algorithm~\ref{cir:rev} details the zero-knowledge proof (ZK) circuit that powers revealing transfers. 
This circuit allows a user to prove that a transfer is valid, meaning it adheres to the defined rules (e.g., ownership, proper balances, etc), while selectively revealing the details of specific output tokens on the blockchain. 
The remaining output tokens remain concealed, preserving privacy.

The Revealing Transfer Witness ($RTW$) data type encapsulates the private data required to construct a valid revealing transfer proof. 
This includes the preimages and paths of input tokens ($inputs$), the sender's secret key ($sk$), and, optionally, the preimages of output tokens ($outputs$). 
The $outputs$ field is only necessary when some of the transaction's output tokens will remain undisclosed.

The Revealing Transfer Public Inputs ($RTPI$) data type encompasses the following public information: 
the lists of nullifiers and grabbers, which represent the consumed tokens; 
the output tokens list ($outputs$), containing the preimages of the tokens to be revealed;
the optional list of commitments ($comms$) for undisclosed output tokens;
the auditor's encrypted copy of hidden data ($audit_d$);
and finally, for the $TK$ contract: its token type ($type_t$), root of the commitment tree ($root_c$), grabber nonce ($nonce_g$), and audit account ($audit_{acc}$). 

The $proveRevealingTransfer$ function (line~\ref{cir.rev_prove_rev}) is the core function responsible for generating the zero-knowledge proof for a revealing transfer. 
It performs the following steps: First, it validates the input tokens using the $checkInputs$ function (defined in Algorithm~\ref{cir:trans_aux}). 
Second, it verifies the correctness of the publicly disclosed output tokens via the $checkClearOutputs$ function. 
Subsequently, it calls $checkMassWithClearOutputs$ to ensure the conservation of fungible and nonfungible assets throughout the transfer. 
The $checkAuditData$ function (defined in Algorithm~\ref{cir:trans_aux_2}) then verifies that the encrypted audit data is consistent with the transaction details. 
Finally, it utilizes the $convertToProof$ function (described in Section~\ref{sec:zk}) to transform the validated witness data into a succinct zero-knowledge proof.

The $checkClearOutputs$ function (line~\ref{cir.rev_check_clear_outputs}) checks the validity of each output token in the $outputs$ list of the public inputs ($RTPI$). 
For each output token image $o$, 
it verifies that either the token ID ($o.id$) or the amount ($o.amount$) is non-zero (ensuring a valid token), 
that the token type ($o.type$) matches the type specified in the public inputs ($pub.type_t$), 
and that the nonce ($o.nonce$) is set to zero, as the nonce is not relevant for revealed outputs.

The $checkMassWithClearOutputs$ function (line~\ref{cir.rev_check_mass}) enforces the conservation principle for both fungible and nonfungible tokens in the transfer. 
It computes the total value of the consumed input tokens and the total value of both the revealed and concealed output tokens. 
It then asserts that these two totals are equal, ensuring that no value is created or destroyed. 
For nonfungible tokens (NFTs), it verifies that the set of consumed NFT IDs exactly matches the set of generated NFT IDs, preventing creation or loss of NFTs.

\circuit{Revealing Transfer}{\label{cir:rev}}{
\INDENT{\textbf{Data Types:}}
    \STATE $RTW:$                                                                           \COMMENT{revealing transfer witness}            \label{data:rev_witness}
        \STATE \tab $inputs: List\langle Img\_Path \rangle,$                                \COMMENT{inputs, alg.~\ref{cir:trans}}          \label{data:rev_wit.inputs}
        \STATE \tab $sk: uint256,$                                                          \COMMENT{payer's secret key}                    \label{data:rev_wit.sk}
        \STATE \tab $[outputs: List\langle TPre\rangle,]$                                   \COMMENT{optional, hidden outputs preimage}    \label{data:rev_wit.outputs} 
    \STATE $RTPI:$                                                                          \COMMENT{revealing transfer's public inputs}    \label{data:rev_public_inputs}
        \STATE \tab $nulls: List\langle TNul \rangle,$                                      \COMMENT{set of nullifiers}                    \label{data:rev_pub.nullifiers}
        \STATE \tab $grabs: List\langle TGrab \rangle,$                                     \COMMENT{set of grabbers}                      \label{data:rev_pub.grabbers}
        \STATE \tab $[comms: List\langle TCom \rangle,]$                                    \COMMENT{optional, hidden outputs}              \label{data:rev_pub.comms} 
        \STATE \tab $outputs: List\langle TPre \rangle,$                                    \COMMENT{clear outputs}                         \label{data:rev_pub.outputs} 
        \STATE \tab $type_t: uint256,$                                                      \COMMENT{token type}                            \label{data:rev_pub.type}  
        \STATE \tab $root_c: MRoot,$                                                        \COMMENT{commitment's tree root}           \label{data:rev_pub.root_c}
        \STATE \tab $nonce_g: uint256,$                                                     \COMMENT{contract's grabber nonce}              \label{data:rev_pub.nonce_g}
        \STATE \tab $audit_d: Bytes,$                                                       \COMMENT{audit data, alg.~\ref{cir:token_nullifier}}\label{data:rev_pub.audit_data}
        \STATE \tab $audit_{acc}: uint256,$                                                 \COMMENT{audit's account}                       \label{data:rev_pub.audit_acc}
\ENDINDENT
\vspace{2.5mm}
\INDENT{\textbf{proveRevealingTransfer}$(wit: RTW, pub: RTPI) : uint256$} \label{cir.rev_prove_rev}
    \require $ checkInputs(wit, pub)$                                   \COMMENT{alg.~\ref{cir:trans_aux}}
    \require $ checkOutputs(wit, pub)$                                  \COMMENT{alg.~\ref{cir:trans_aux}}
    \require $ checkClearOutputs(pub)$
    \require $ checkMassWithClearOutputs(wit, pub)$    
    \require $ checkAuditData(wit, pub)$                                \COMMENT{alg.~\ref{cir:trans_aux_2}}
    \RETURN  $ convertToProof(wit)$                                     \COMMENT{section~\ref{sec:zk}}
\ENDINDENT

\vspace{2.5mm}
\INDENT{\textbf{checkClearOutputs}$(pub: RTPI) : bool$}\label{cir.rev_check_clear_outputs}
    \FORALL{$o \in pub.outputs$}
        \require $ o.id \ne 0 \lor o.amount \ne 0$      \COMMENT{valid token}
        \require $ o.type \eq pub.type_t$               \COMMENT{correct type}      
        \require $ o.nonce \eq 0$                       \COMMENT{nonce is irrelevant}      
    \ENDFOR
    \RETURN \TRUE
\ENDINDENT

\vspace{2.5mm}
\INDENT{\textbf{checkMassWithClearOutputs}$(wit: RTW, pub: RTPI) : bool$} \label{cir.rev_check_mass}
    \STATE   $ imgs_{in} \get \{i.img,\;\forall i \in wit.inputs\}$                                              
    \STATE   $ total_{in} \get getAmountSum(imgs_{in})$                             
    \STATE   $ hidden_{out} \get getAmountSum(wit.outputs)$                         
    \STATE   $ exposed_{out} \get getAmountSum(pub.outputs)$                          
    \STATE   $ total_{out} \get hidden_{out} + exposed_{out}$                          
    \require $ total_{in} \eq total_{out}$ \COMMENT{fungible mass}
    \STATE   $ input_{ids} \get \{i.id,\;\forall i \in imgs_{in}, i.id \ne 0\}$                  
    \STATE   $ hidden_{ids} \get \{o.id,\;\forall o \in wit.outputs, o.id \ne 0\}$              
    \STATE   $ exposed_{ids} \get \{o.id,\;\forall o \in pub.outputs, o.id \ne 0\}$               
    \STATE   $ output_{ids} \get hidden_{ids} \cup exposed_{ids}$ 
    \require $ input_{ids} \eq output_{ids}$ \COMMENT{nonfungible mass}
    \RETURN \TRUE
\ENDINDENT
}

Algorithm~\ref{alg:rev_trans} defines the logic for the Revealing Transfer Flow within the Token smart contract. 
It outlines the data types, functions, and execution flow for processing revealing transfer transactions. 
The $RevealingTransferTransaction$ (RTT) structure represents a revealing transfer transaction. 
It consists of two fields: $pub$, which holds the public inputs required for verification (RTPI), and $proof$, which stores the zero-knowledge proof associated with the transaction.
The contract flow includes two functions to handle revealing transfers. 
The $revealingTransfer$ public function orchestrates the execution of a revealing transfer. 
It first verifies the transaction using the correspondent verifier, ensuring the validity of the provided proof and public inputs. 
If verification succeeds, it calls the $doRevealingTransfer$ function to perform the transfer. 
Finally, it emits events to signal completion.

\contract{Revealing Transfer Flow}{\label{alg:rev_trans}}{
\INDENT{\textbf{Data Types:}}       
    \STATE $RTT: $                                                  \COMMENT{revealing transfer's transaction}                  \label{data:rev_trans}
        \STATE \tab $pub: RTPI$                                     \COMMENT{revealing transfer's public input}                 \label{data:rev_trans.pub}  
        \STATE \tab $proof: uint256$                                \COMMENT{zero-knowledge proof}                              \label{data:rev_trans.proof}
\ENDINDENT

\vspace{2.5mm}    
\INDENT{\textbf{+ revealingTransfer}$(t: RTT)$}     \label{alg.rev}
    \require $revealing_v.verify(t)$                \label{alg.rev.req.verify}
    \STATE   $doRevealingTransfer(t.pub)$           \label{alg.rev.do}  
    \events                                         \label{alg.rev.events}  
\ENDINDENT

\vspace{2.5mm}    
\INDENT{\textbf{- doRevealingTransfer}$(pub: RTPI)$}                            \label{alg.dorev} 
    \require $ pub.audit_{acc} \eq this.audit\_acc$                             \label{alg.dorev.req.audit_pk}
    \require $ pub.type_t \eq this.type_t$                                      \label{alg.dorev.req.type}
    \require $ pub.nonce_g \eq this.grab\_nonce$                                \label{alg.dorev.req.grabb_nonce}
    \require $ pub.root_c \in this.tree_c.roots$                                \label{alg.dorev.req.root_c}         
    \require $ \forall g \in pub.grabs,\ g \notin this.grabbers$             \label{alg.dorev.req.grabbers}
    \require $ \forall n \in pub.nulls,\  n \notin this.nullifiers$        \label{alg.dorev.req.nullifiers}
    \require $ \forall c \in pub.comms,\  c \notin this.tree_c$                 \label{alg.dorev.req.comms}
    \require $ \forall o \in pub.outputs,\  o.owner \is SmartContract$          \label{alg.dorev.req.owner}    
    \STATE   $ this.nullifiers \cupeq pub.nulls$                           \label{alg.dorev.add.nullifiers}
    \STATE   $ this.grabbers \cupeq pub.grabs$                               \label{alg.dorev.add.grabbers}
    \STATE   $ this.tree_c \cupeq pub.comms$                                    \label{alg.dorev.add.commitments}
    \FORALL{$out \in pub.outputs$}        
        \STATE $balances[out.owner] \peq out.amount$                
        \STATE $nfts[out.owner] \cupeq out.id$                
    \ENDFOR
\ENDINDENT
}

The $doRevealingTransfer$ function (line~\ref{alg.dorev}) handles the core logic of the revealing transfer. 
It starts by checking several conditions to ensure the transaction's validity. 
This includes verifying the audit account ($pub.audit_{acc}$), token type ($pub.type_t$), grabber nonce ($pub.nonce_g$), commitment root ($pub.root_c$), and ensuring that the provided grabbers ($pub.grabs$) and nullifiers ($pub.nulls$) have not been used before. 
It also requires that all output owners ($pub.outputs$) are smart contracts. 
If all conditions are met, the function updates the contract's state by adding the provided nullifiers and grabbers to their respective sets. 
Then, it iterates through the outputs ($pub.outputs$) and updates the balances or NFTs accordingly. 

\subsubsection{Hiding Transfer Transaction}

In contrast to revealing transfers, which typically expose assets on the blockchain, hiding transfers provide a mechanism for transferring assets while simultaneously concealing the sender's and the new owner's identities. 
Hiding transfers achieve this by consuming exposed fungible and/or nonfungible assets (e.g. made visible through a previous revealing transfers) and creating their ``hidden" version on the blockchain. 
While the consumed amount and type are indirectly discernible from the public inputs, the key privacy enhancement comes from obscuring the new owner. 

\begin{figure}[!ht]
    \centering
    \includegraphics[width=1\linewidth]{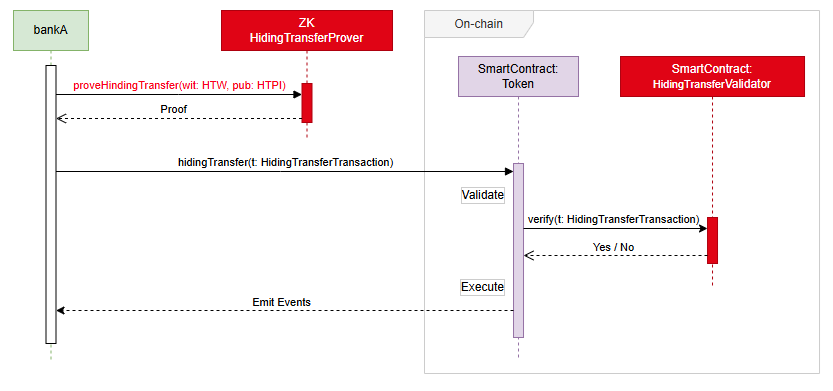}
    \caption{Hiding transfer flow}
    \label{fig:hiding.transfer.flow}
\end{figure}

Figure~\ref{fig:hiding.transfer.flow} depicts the hiding transfer flow, analogous to the revealing transfer described in Section~\ref{sec:revealing_transfer}. 
The core distinction lies in its purpose: concealing transaction output details. 
This is achieved through components tailored for hiding transfers.
Specifically, the ZK $HidingTransferProver$ circuit and $HidingTransferValidator$ smart contract replace their revealing counterparts. 
These components process $HTW$ (Hiding Transfer Witness) and $HTPI$ (Hiding Transfer Public Input), data structures designed for hiding transfers. 
Similarly, the functions $proveHidingTransfer$ (in the prover) and $hidingTransfer$ (in the token contract) reflect the hiding nature of this process. 

\noindent \textit{Note: To enhance readability, this section omits the fields, data types, and functions related to token burning. These features can be implemented by adapting the mechanisms described in Section~\ref{sec:transfer_burn}.}

Algorithm~\ref{cir:hid} presents the Hiding Transfer ZK circuit, which facilitates the transfer of assets while concealing the sender's identity and the transferred amount.
The Hiding Transfer Witness ($HTW$) data type  encapsulates the private data required for a hiding transfer. 
This includes a list of output tokens ($outputs$) being created and optionally the consumed tokens' owner's secret key ($sk$).

The Hiding Transfer Public Inputs ($HTPI$) data type comprises the information that will be publicly available to the circuit verifier on the blockchain. 
It includes: 
the consumed fungible amount ($amount_i$) and/or the consumed NFT IDs ($ids_i$); 
the sender's public account ($acc_i$) if available; 
commitments for the newly generated output tokens; 
the audit data ($audit_d$); 
the $TK$ contract token type ($type_t$) and auditor's account address ($audit_{acc}$).

The $proveHidingTransfer$ function is the main entry point for generating a zero-knowledge proof for a hiding transfer. 
It executes several checks to ensure the transfer's validity: 
First, it ensures that either a fungible amount or a set of NFT IDs are being consumed. 
Second, it verifies that if the consumed token's owner account ($acc_i$) is provided, and it correctly corresponds to the sender's secret key ($sk$). 
Then, it invokes auxiliary functions to validate the output tokens, to ensure the conservation of fungible and nonfungible assets (i.e. that the total input value equals the total output value), and to validate the audit data. 
Finally, the function converts the witness data into a zk-SNARKS using the $convertToProof$ symbolic function. 

The $checkMassWithClearInputs$ function (line~\ref{cir.hid_check_mass_clear_inputs}) enforces the conservation principle for both fungible and nonfungible tokens within the context of a hiding transfer. 
Similar to the $checkMass...$ functions used in other transfer types, it ensures that no value is created or destroyed during the transfer. 
However, $checkMassWithClearInputs$ is unique in that it verifies the conservation by comparing the publicly visible consumed assets (inputs of the hiding transfer) with the privately generated output tokens (outputs of the hiding transfer).

\circuit{Hiding Transfer}{\label{cir:hid}}{
\INDENT{\textbf{Data Types:}}       
    \STATE $HTW:$                                                   \COMMENT{transfer's witness}                    \label{data:hid_witness}
        \STATE \tab $outputs: List\langle TPre \rangle,$            \COMMENT{list of outputs}                       \label{data:hid_wit_outputs} 
        \STATE \tab $[sk: SKey]$                                    \COMMENT{optional, owner's secret key}          \label{data:hid_wit_sk}
    \STATE $HTPI:$                                                  \COMMENT{transfer's public inputs}              \label{data:hid_public_inputs}
        \STATE \tab $[amount_i: uint256,]$                          \COMMENT{optional, fungible input}              \label{data:hid_pub_amount}
        \STATE \tab $[ids_i: List\langle uint256 \rangle,]$         \COMMENT{optional, nonfungible input}          \label{data:hid_pub_ids}
        \STATE \tab $[acc_i:  Account,]$                            \COMMENT{optional, public account}              \label{data:hid_pub_account}
        \STATE \tab $comms: List\langle TCom \rangle,$              \COMMENT{commitments}                           \label{data:hid_pub_commitments}  
        \STATE \tab $type_t: uint256,$                              \COMMENT{token type}                            \label{data:hid_pub_type}  
        \STATE \tab $audit_d: Bytes,$                               \COMMENT{audit data, alg.~\ref{cir:token_nullifier}}\label{data:hid_pub_audit_data}
        \STATE \tab $audit_{acc}: Account,$                         \COMMENT{audit account}                         \label{data:hid_pub_auth_pk}
\ENDINDENT

\vspace{2.5mm}
\INDENT{\textbf{proveHidingTransfer}$(wit: HTW, pub: HTPI) : uint256$} \label{cir.hid_prove_hid_transfer}
    \require $ pub.amount_i \ne 0 \lor pub.ids_i \ne 0$    
    \require $ pub.acc_i \eq 0 \lor pub.acc_i \eq getAccount(wit.sk)$
    \require $ checkOutputs(wit, pub)$                                      \COMMENT{alg.~\ref{cir:trans_aux}}
    \require $ checkMassWithClearInputs(wit, pub)$    
    \require $ checkAuditData(wit, pub)$                                    \COMMENT{alg.~\ref{cir:trans_aux_2}}
    \RETURN  $ convertToProof(wit)$                                         \COMMENT{section~\ref{sec:zk}}
\ENDINDENT

\vspace{2.5mm}
\INDENT{\textbf{checkMassWithClearInputs}$(wit: HTW, pub: HTPI) : bool$}  \label{cir.hid_check_mass_clear_inputs}
    \STATE   $ total_{out} \get sum(\{o.amount,\;\forall o \in wit.outputs\})$    \label{cir.hid_check_mass_amount}                      
    \require $ pub.amount_i \eq total_{out}$
    \STATE   $ output_{ids} \get \{o.id,\;\forall o \in wit.outputs \land o.id \ne 0\}$ \label{cir.hid_check_mass_ids}
    \require $ pub.ids_i \eq output_{ids}$
    \RETURN \TRUE
\ENDINDENT
}

Algorithm~\ref{alg:hid_trans} outlines the data type and functions related to hiding transfers in the $TK$ smart contract. 
These functions are designed to work in conjunction with zero-knowledge proofs (ZKPs) circuit detailed in this same section to ensure the privacy and security of hiding operations, allowing users to conceal the details of their transactions' outputs while still maintaining the integrity of the system.

The algorithm starts defining the Hiding Transfer Transaction data type ($HTT$), which includes the Hiding Transfer Public Inputs ($HPI$) and a $proof$ (a zk-SNARK for validation).
The core functionality is provided by the $hidingTransfer$ function. 
This public function processes incoming hiding transfer transactions. 
It first calls a hidden verifier to check the ZKP provided in the transaction, ensuring its validity (line~\ref{alg.hid_req_verify}). 
If the proof is valid, the $doHidingTransfer$ function is invoked to execute the hiding transfer (line~\ref{alg.hid_do}).

The $doHidingTransfer$ function performs the necessary checks and state updates to complete the hiding transfer. 
It ensures that the token type ($pub.type\_t$) in the transaction matches the contract's token type (line~\ref{alg.dohid_req_type}) and that the commitments in $pub.comms$ are not already present in the $TK$'s commitment tree $this.tree\_c$ (line~\ref{alg.dohid_req_tree}). 
It also verifies that the informed audit account ($pub.audit_{acc}$) matches the contract's audit account (line~\ref{alg.dohid_req_audit_acc}). 
The function then determines the owner of the tokens to be hidden, using $pub.acc\_i$ if it's not zero, otherwise defaulting to the message sender ($msg.sender$) (line~\ref{alg.dohid_owner}). 
It checks if the owner has a sufficient balance of fungible tokens (line~\ref{alg.dohid_req_amount}) and possesses all the specified nonfungible tokens (NFTs) (line~\ref{alg.dohid_req_ids}). 
Subsequently, it deducts the specified amount ($pub.amount\_i$) from the owner's public fungible token balance (line~\ref{alg.dohid_sub_amount}) and removes the specified NFTs ($pub.ids\_i$) from the owner's public NFT set (line~\ref{alg.dohid_sub_ids}). 
Finally, it adds the new commitments to the commitment tree, effectively hiding the transferred tokens (line~\ref{alg.dohid_add_comms}).

\algoritmo{Hiding Transfer Data Type and Flow}{\label{alg:hid_trans}}{
\INDENT{\textbf{Data Types:}}  
    \STATE $HTT: $                                      \COMMENT{hiding transfer's transaction}       \label{data:hid_trans}
        \STATE \tab $ pub: HPI$                         \COMMENT{hiding transfer's public data}       \label{data:hid_trans_transfer} 
        \STATE \tab $ proof: uint256$                   \COMMENT{ZK proof}                            \label{data:hid_trans_proof}  
\ENDINDENT

\vspace{2.5mm}   
\INDENT{\textbf{+ hidingTransfer}$(t: HTT)$  \label{alg.hid}}
    \require $ hidden_v.verify(t)$                                                                 \label{alg.hid_req_verify}
    \STATE   $ doHidingTransfer(t.pub)$                                                            \label{alg.hid_do}  
    \events                                                                                        \label{alg.hid_events}  
\ENDINDENT
\vspace{2.5mm}   
\INDENT{\textbf{- doHidingTransfer}$(pub: HTPI)$                        \label{alg.dohid}}
    \require $ pub.type_t \eq this.type_t$                              \label{alg.dohid_req_type}
    \require $ pub.comms \notin this.tree_c$                            \label{alg.dohid_req_tree}
    \require $ pub.audit_{acc} \eq this.audit\_acc$                     \label{alg.dohid_req_audit_acc}
    \STATE   $ owner \get (pub.acc_i \ne 0)?\; pub.acc_i:msg.sender $   \label{alg.dohid_owner}
    \require $ this.balances[owner] \geq pub.amount_i$                  \label{alg.dohid_req_amount}              
    \require $ this.nfts[owner] \supseteq pub.ids_i$                    \label{alg.dohid_req_ids}  
    \STATE   $ this.balances[owner] \meq pub.amount_i$   \COMMENT{consume amount}   \label{alg.dohid_sub_amount}                       
    \STATE   $ this.nfts[owner] \capeq pub.ids_i$        \COMMENT{consume NFTs}     \label{alg.dohid_sub_ids}                
    \STATE   $ this.tree_c \cupeq pub.comms$             \COMMENT{add new tokens}   \label{alg.dohid_add_comms}                    
\ENDINDENT
}

\subsubsection{Grab Transaction}

The Grab circuit, as detailed in Algorithm~\ref{cir:grab}, outlines the procedure for seizing tokens, typically initiated by an authority figure due to legal or regulatory reasons. 
This process employs zero-knowledge proofs to ensure the validity of the seizure while preserving the privacy of the involved parties.

The Grabber Witness $GW$ data type defines the structure of the witness, which includes a list of input tokens ($inputs$), a list of output tokens ($outputs$), the authority's secret key ($auth_{sk}$), the owner's public key ($owner_{pk}$), and the owner's grabber key ($grabber_k$). 
The Grabber Public Inputs $GPI$ data type encompasses the public inputs, including a list of grabbers ($grabs$), a list of commitments ($comms$), the contract's token type ($type_t$), the root of the commitment tree ($root_c$), a grabber nonce ($nonce_g$), and the authority's account ($auth_{acc}$).

The $proveGrabber$ function orchestrates the proof generation process. 
It first verifies that the provided authority account $pub.auth_{acc}$ is derived from the provided authority secret key $wit.auth_{sk}$. 
Then it invokes the $checkGrabbInputs$ function to validate the inputs, ensuring they meet the necessary criteria. 
It then utilizes auxiliary functions, namely $checkOutputs$ (from Algorithm $\ref{cir:trans_aux}$), $checkMassConservation$ (from Algorithm $\ref{cir:trans_aux_2}$), to verify the outputs and ensure mass conservation respectively. 
Finally, it converts the witness into a zero-knowledge proof using the $convertToProof$ symbolic function.

The $checkGrabInputs$ function verifies that there are no duplicate input images within the $wit.inputs$ list (line~\ref{cir.check_grab_req_img}). 
It ensures that the number of provided public grabbers matches the number of inputs (line~\ref{cir.check_grab_req_size}). 
It also verifies that the provided nonce ($pub.nonce\_g$) matches the result of decrypting the provided owner's grabber key ($wit.grabber\_k$) with the provided owner's public key ($wit.owner_{pk}$) (line~\ref{cir.check_grab_req_nonce}). 
For each input (line~\ref{cir.check_grab_for}), it extracts the image ($img$) (line~\ref{cir.check_grab_get_img}) and path ($path$) (line~\ref{cir.check_grab_get_path}), calculates the grabber ($grab$) (line~\ref{cir.check_grab_get_grab}) and the root of the commitment tree ($root$) (line~\ref{cir.check_grab_get_root}), and then performs the following checks: 
the image must have either a non-zero amount or a non-zero ID (line~\ref{cir.check_grab_req_not_zero}); 
the token type of the image must match the provided public token type (line~\ref{cir.check_grab_req_type}); 
the calculated grabber must match the corresponding public grabber (line~\ref{cir.check_grab_req_grab}); 
and the calculated root must match the provided public root (line~\ref{cir.check_grab_req_root}). 
The function returns true if all these conditions are met.

\circuit{Grab}{\label{cir:grab}}{
\INDENT{\textbf{Data Types:}}
    \STATE $GW:$                                                        \COMMENT{grabber's witness}                     \label{data:grab_witness}
        \STATE \tab $inputs: List\langle Img\_Path \rangle,$            \COMMENT{inputs, algorithm~\ref{cir:trans}}     \label{data:grab_wit_inputs}
        \STATE \tab $outputs: List\langle TPre \rangle,$                \COMMENT{list of outputs}                       \label{data:grab_wit_outputs} 
        \STATE \tab $auth_{sk}: SKey,$                                  \COMMENT{authority's secret key}                \label{data:grab_wit_auth_sk}
        \STATE \tab $owner_{pk}: PKey,$                                 \COMMENT{owner's public key}                    \label{data:grab_wit_owner_pk}
        \STATE \tab $grabber_k: GKey,$                                  \COMMENT{owner's grabber key}                   \label{data:grab_wit_grabber_k}
    \STATE $GPI:$                                                       \COMMENT{grabber's public inputs}               \label{data:grab_public_inputs}
        \STATE \tab $grabs: List\langle TGra \rangle,$                  \COMMENT{grabbers}                              \label{data:grab_pub_grabbers}
        \STATE \tab $comms: List\langle TCom \rangle,$                  \COMMENT{commitments}                           \label{data:grab_pub_commitments}
        \STATE \tab $type_t: Type,$                                     \COMMENT{token type}                            \label{data:grab_pub_type}  
        \STATE \tab $root_c: MRoot,$                                    \COMMENT{root of the commitment tree}           \label{data:grab_pub_root_c}
        \STATE \tab $nonce_g: Nonce,$                                   \COMMENT{grabber nonce in the contract}         \label{data:grab_pub_grab_nonce}
        \STATE \tab $auth_{acc}: Account$                               \COMMENT{authority's account}                   \label{data:grab_pub_auth}
\ENDINDENT

\vspace{2.5mm}
\INDENT{\textbf{proveGrabber}$(wit: GW, pub: GPI) : uint256$}                                             
    \require $ pub.aut_{acc} \eq getAccount(wit.auth_{sk})$  \COMMENT{owns $auth\_acc$}
    \require $ checkGrabbInputs(wit, pub)$  
    \require $ checkOutputs(wit, pub)$                       \COMMENT{alg.~\ref{cir:trans_aux}}
    \require $ checkMassConservation(wit)$                   \COMMENT{alg.~\ref{cir:trans_aux_2}}
    \RETURN  $ convertToProof(wit)$
\ENDINDENT

\vspace{2.5mm}
\INDENT{\textbf{checkGrabInputs}$(wit: TW, pub: TPI): bool$}  \label{cir.check_grabb_inputs}
    \require $ \nexists\; a,b \in wit.inputs: a.img \eq b.img$           \label{cir.check_grab_req_img}            
    \require $ pub.grabs.size \eq wit.inputs.size$                       \label{cir.check_grab_req_size}      
    \require $ pub.nonce_g \eq wit.owner_{pk}.decypher(wit.grabber_k)$   \label{cir.check_grab_req_nonce}        
    \FOR{$i\get 0 \;to\; wit.inputs.size$}                               \label{cir.check_grab_for}                         
        \STATE   $ img \get wit.inputs[i].img$                           \label{cir.check_grab_get_img}                            
        \STATE   $ path \get wit.inputs[i].path$                         \label{cir.check_grab_get_path}                              
        \STATE   $ grab \get grab(img, wit.grabber_k)$                   \label{cir.check_grab_get_grab}                           
        \STATE   $ root \get getRoot(commit(img), path)$                 \label{cir.check_grab_get_root}            
        \require $ img.amount \ne 0 \lor img.id \ne 0$                   \label{cir.check_grab_req_not_zero}          
        \require $ pub.type_t \eq img.type$                              \label{cir.check_grab_req_type}                   
        \require $ pub.grabs[i] \eq grab$                                \label{cir.check_grab_req_grab}  
        \require $ pub.root_c \eq root$                                  \label{cir.check_grab_req_root}       
    \ENDFOR
    \RETURN \TRUE
\ENDINDENT
}

Algorithm~\ref{alg:grab} outlines the data types and functions related to the "grab" functionality within the $TK$ smart contract. 
This feature, guarded by a designated authority, allows for the retrieval of specific tokens from the commitment tree, effectively taking the possession and property of a participant's tokens under controlled circumstances.

The algorithm initiates defining the Grab Transaction data type ($GT$), which includes the Grab Public Inputs ($GPI$) and a $proof$, a zero-knowledge proof used for validation.
The core functionality is provided by the $grab$ function, a public function that can only be executed by the designated authority address ($this.auth\_add$) (line~\ref{alg.grab_req_auth_add}). 
It ensures that the provided authority account ($t.auth_{acc}$) matches the contract's stored authority account (line~\ref{alg.grab_req_auth_acc}). 
The function first calls the grabber verifier, $grabber\_v.verify$, to validate the provided zero-knowledge proof (line~\ref{alg.grab_req_verify}). 
If the proof is valid, the $doGrab$ function is invoked (line~\ref{alg.grab_do}).

The $doGrab$ function performs the necessary checks and updates to execute the grab operation. 
It verifies that the provided token type ($pub.type\_t$) matches the contract's token type (line~\ref{alg.dograb_req_type}). 
It also checks that the provided commitment tree root ($pub.root\_c$) is a valid root in the contract's commitment tree roots set (line~\ref{alg.dograb_req_root}). 
It ensures that none of the provided grabbers ($pub.grabs$) are already present in the contract's set of used grabbers (line~\ref{alg.dograb_req_grabbers}) and that none of the provided commitments ($pub.comms$) are already present in the $TK$'s commitment tree (line~\ref{alg.dograb_req_commitments}). 
If these conditions are met, the function adds the provided grabbers to the contract's set of used grabbers (line~\ref{alg.dograb_add_grabbers}) and adds the provided commitments to the contract's commitment tree (line~\ref{alg.dograb_add_commitments}). 

\contract{Grab}{\label{alg:grab}}{
\INDENT{\textbf{Data Types:}}  
    \STATE $GT:$                                    \COMMENT{grabber's transaction}     \label{data:grab}
        \STATE \tab $pub: GPI,$                     \COMMENT{grabber's public inputs}   \label{data:grab_grabber}
        \STATE \tab $proof: uint256$                \COMMENT{zero-knowledge proof}      \label{data:grab_proof}
\ENDINDENT
\vspace{2.5mm}    
\INDENT{\textbf{+ grab}$(t: GT)$} \label{alg.grab}
    \require $ msg.sender \eq this.auth\_add$       \COMMENT{correct authority EOA}     \label{alg.grab_req_auth_add}
    \require $ t.auth_{acc} \eq this.auth\_acc$     \COMMENT{correct authority account} \label{alg.grab_req_auth_acc}            
    \require $ grabber_v.verify(t)$                 \COMMENT{verify proof}              \label{alg.grab_req_verify}    
    \STATE   $ doGrab(t.pub)$                                                           \label{alg.grab_do}    
    \events                                                                             \label{alg.grab_events}  
\ENDINDENT
\vspace{2.5mm}
\INDENT{\textbf{+ doGrab}$(pub: GPI)$} \label{alg.dograb}
    \require $ pub.type_t \eq this.type_t$                                              \label{alg.dograb_req_type}
    \require $ pub.root_c \in this.tree_c.roots$                                        \label{alg.dograb_req_root}       
    \require $\forall g \in pub.grabs, g \notin this.grabbers$                          \label{alg.dograb_req_grabbers}
    \require $\forall c \in pub.comms, c \notin this.tree_c$                            \label{alg.dograb_req_commitments}
    \STATE   $\forall g \in pub.grabs, this.grabbers[g] \get \TRUE$                     \label{alg.dograb_add_grabbers}        
    \STATE   $this.tree_c \cupeq pub.comms$                                             \label{alg.dograb_add_commitments}
\ENDINDENT
}

\subsubsection{Delegated Mint Transaction}

\begin{figure}[!ht]
    \centering
    \includegraphics[width=1\linewidth]{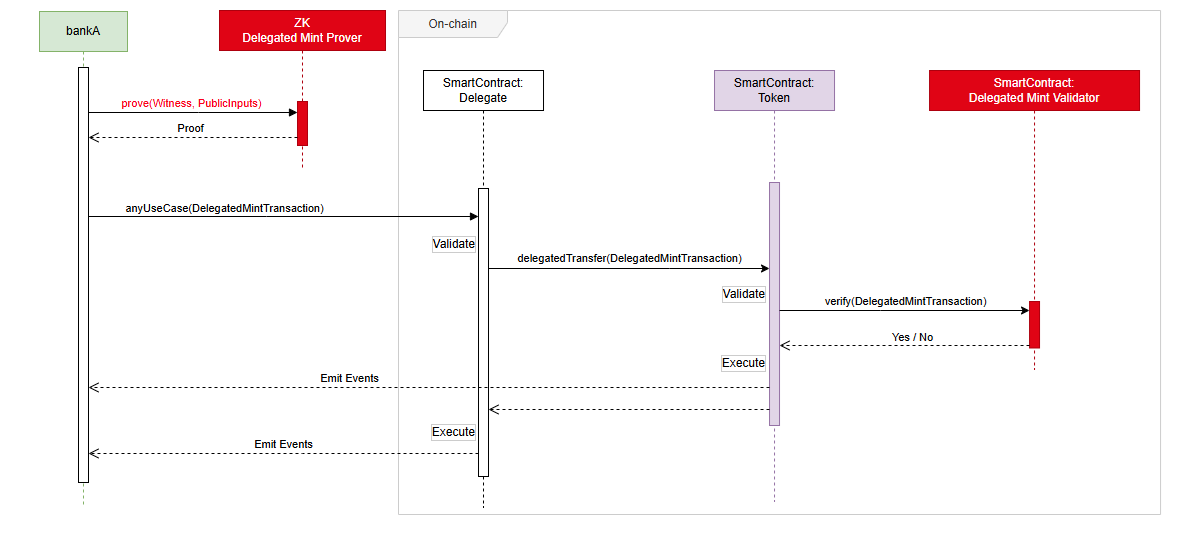}
    \caption{Delegated Mint Flow}
    \label{fig:del_mint_flow}
\end{figure}

Algorithm~\ref{cir:del_mint} describes the zero-knowledge proof (ZK) circuit for delegated minting, where a designated contract is authorized to mint tokens on behalf of another entity. 
This mechanism allows for greater flexibility and efficiency in token issuance while maintaining security and privacy.

The $DelegatedMintWitness$ data type extends the $MintWitness$ and includes the necessary information for the delegated minting process. 
The $DelegatedMintPublicInputs$ data type encompasses the public inputs, including the standard $MintPublicInputs$, the address of the delegated contract ($del_{add}$), a binding value for the delegated contract ($del_b$), and the ZK proof.

The core of the algorithm lies in the $proveDelegatedMint$ function. 
This function takes the $DelegatedMintWitness$ and $DelegatedMintPublicInputs$ as inputs and returns a ZK proof. 
It first invokes the $proveMint$ function (from Algorithm~\ref{cir:mint}) to generate a proof based on the standard minting witness and public inputs. 
This ensures that the underlying minting operation adheres to the established rules.

Next, the function calculates a hash of the witness ($hash_{wit}$) using the $hash256$ function. 
This hash serves as a unique identifier for the witness data. 
The function then requires that the delegated contract binding ($del_b$) matches the hash of the delegated contract address and the witness hash. 
This requirement ensures that the proof is linked to the specific delegated contract and witness, preventing unauthorized use of the proof.

Finally, the function converts the witness into a ZK proof using the $convertToProof$ function and returns this proof. 
This proof can be used to verify the validity of the delegated minting operation without revealing the private information contained in the witness. 

\algoritmo{Delegated Mint's Types and ZK Circuit}{\label{cir:del_mint}}{
\STATE $DMW \is MW$                             \COMMENT{delegated mint's witness}              \label{data:del_mint_witness}
\STATE $DMPI:$                                  \COMMENT{delegated mint's public inputs}        \label{data:del_mint_pub}
    \STATE \tab $pub: MPI,$                     \COMMENT{mint's public inputs}                  \label{data:del_mint_pub_pub}
    \STATE \tab $del_{add}: address$            \COMMENT{delegated contract address}            \label{data:del_mint_pub_del_add} 
    \STATE \tab $del_b: uint256$                \COMMENT{delegated contract binding}            \label{data:del_mint_pub_del_bind} 
    \STATE \tab $proof: uint256$                \COMMENT{ZK proof}                              \label{data:del_mint_pub_proof} 
\vspace{2.5mm}
\INDENT{\textbf{proveDelegatedMint}$(wit: DMW, pub: DMPI): uint256$}
    \require $ checkDelegate(wit, pub)$
    \STATE   $ proveMint(wit, pub.pub)$                         \COMMENT{alg.~\ref{cir:mint}}
    \RETURN  $ convertToProof(wit)$
\ENDINDENT
\vspace{2.5mm}
\INDENT{\textbf{checkDelegate}$(wit: DMW, pub: DMPI): bool$}                                    \label{cir.check_delegate}
    \STATE  $ hash_{wit} \get hash256(wit)$
    \RETURN $ pub.del_b \eq hash256(pub.del_{add}, hash_{wit})$
\ENDINDENT
}

This code describes the Delegated Mint Smart Contract Flow within the Token smart contract, outlining the process of minting new tokens through delegation.

It first defines the $DelegatedMintTransaction$ structure, which represents a delegated minting transaction. 
This structure includes $pub$ for holding the public inputs required for verification and $proof$ for storing the zero-knowledge proof (ZKP) associated with the transaction.

The core logic of the delegated minting process is encapsulated within the $delegatedMint$ public function. 
This function enforces several requirements before proceeding with minting. 
Firstly, it checks if the transaction sender ($msg.sender$) matches the delegate address ($t.del_{add}$) specified in the transaction. 
Secondly, it verifies that the sender is either a registered issuer in the $issuers$ mapping or that a valid issuer root ($t.pub.root_i$) is provided. 
Lastly, it calls the respective verifier function to verify the zero-knowledge proof associated with the transaction.

If all requirements are met, the function calls the $doMint$ function (Algorithm~\ref{alg:mint}, line~\ref{alg.domint}) to execute the actual minting process using the public inputs from the transaction. 
Finally, it emits events to signal the successful completion of the delegated minting operation.

\algoritmo{Delegated Mint's Smart Contract Flow}{\label{alg:del_mint}}{
\STATE $DMT:$                           \COMMENT{delegated mint's transaction}      \label{data:del_mint}     
    \STATE \tab $pub: DMPI,$            \COMMENT{delegated mint's public inputs}    \label{data:del_mint.pub}
    \STATE \tab $proof: uint256$        \COMMENT{ZKP}                               \label{data:del_mint.proof}
\vspace{2.5mm}    
\INDENT{\textbf{+ delegatedMint}$(t: DMT)$}
    \require $ msg.sender \eq t.del_{add}$                                  \COMMENT{caller is the delegate}
    \require $ msg.sender \in this.issuers \lor t.pub.pub.root_i \ne 0$     \COMMENT{caller can mint}
    \require $ del\_mint_v.verify(t)$                                       \COMMENT{verify proof}              
    \STATE   $ doMint(t.pub)$                                               \COMMENT{alg.~\ref{alg:mint}} 
    \events        
\ENDINDENT
}


\subsubsection{Delegated Transfer/Burn Transaction}

This code details the mechanisms for delegated transfers and burns within the Token smart contract, encompassing the witness, public inputs, zero-knowledge circuit, transaction structure, and smart contract logic.

First, it defines the structure for a delegated transfer witness ($DelegatedTransferWitness$), which is equivalent to a standard $TransferWitness$, and the structure for delegated transfer public inputs ($DelegatedTransferPublicInputs$). 
The latter includes the standard transfer public inputs ($pub$), the delegate's address ($del_{add}$), a binding value for the delegate contract ($del_b$), and the zero-knowledge proof ($proof$).

\begin{figure}[!ht]
    \centering
    \includegraphics[width=1\linewidth]{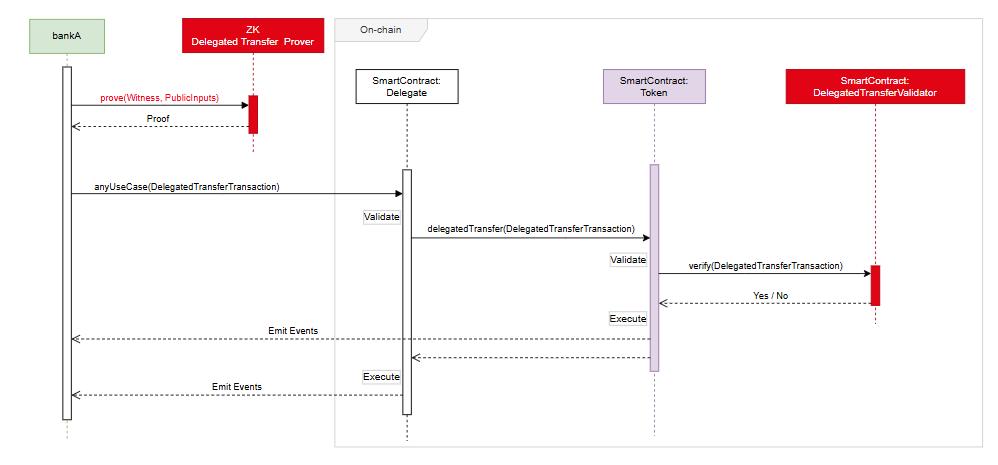}
    \caption{Delegated Transfer/Burn Flow}
    \label{fig:del_transfer_flow}
\end{figure}

The $proveDelegatedTransfer$ function outlines the process of generating a proof for a delegated transfer. 
It first generates a proof for the underlying transfer using $proveTransfer$. 
Then, it hashes the witness and checks if the provided delegate binding ($pub.del_b$) matches the hash of the delegate address and the hashed witness. 
If the check passes, it converts the witness into a proof and returns it.

\algoritmo{Delegated Transfer's Types and ZK Circuit}{\label{cir:del_trans}}{
\STATE $DTW \is TW$                                                 \COMMENT{delegated transfer's witness, alg.~\ref{cir:trans}}    \label{data:del_trans_witness}
\STATE $DTPI:$                                                      \COMMENT{delegated transfer's public inputs}                    \label{data:del_trans_pub}
    \STATE \tab $pub: TPI,$                                         \COMMENT{transfer's public inputs, alg.~\ref{cir:trans}}        \label{data:del_trans_pub.pub}
    \STATE \tab $del_{add}: address$                                \COMMENT{delegated contract address}                            \label{data:del_trans_pub.del_add}  
    \STATE \tab $del_b: uint256$                                    \COMMENT{delegated contract binding}                            \label{data:del_trans_pub.del_bind}  
    \STATE \tab $proof: uint256$                                    \COMMENT{ZK proof}                                              \label{data:del_trans_pub.proof}  
\vspace{2.5mm}
\INDENT{\textbf{proveDelegatedTransfer}$(wit: DTW, pub: DTPI): uint256$}
    \require $ checkDelegate(wit, pub)$         \COMMENT{alg.~\ref{cir:del_mint}}
    \STATE   $ proveTransfer(wit, pub.pub)$
    \RETURN  $ convertToProof(wit)$
\ENDINDENT
}

Next, the code at Algorithm~\ref{alg:del_trans} defines the $DelegatedTransferTransaction$ structure, which encapsulates the public inputs and proof for a delegated transfer/burn. 
The contract state includes $del\_transfer_v: DelegatedTransferVerifier$, which stores the address of the delegated transfer/burn verifier contract.

The $delegatedTransfer$ function handles the execution of a delegated transfer. 
It requires that the transaction sender ($msg.sender$) matches the delegate address ($t.del_{add}$) and that the proof verifies successfully using the delegated transfer verifier. 
If both conditions are met, it calls the $doTransfer$ function (Algorithm~\ref{alg:trans} line~\ref{alg.dotrans}) to execute the transfer and emits events to signal completion.

\algoritmo{Delegated Transfer's Smart Contract Flow}{\label{alg:del_trans}}{
\STATE $DTT:$                       \COMMENT{delegated transfer's transaction}      \label{data:del_trans}     
    \STATE \tab $pub: DTPI,$        \COMMENT{delegated transfer's public inputs}    \label{data:del_trans.pub}
    \STATE \tab $proof: uint256$    \COMMENT{ZKP}                                   \label{data:del_trans.proof}
\vspace{2.5mm}    
\INDENT{\textbf{+ delegatedTransfer}$(t: DTT)$}
    \require $ msg.sender \eq t.del_{add}$                      \COMMENT{caller is the delegate}
    \require $ del\_transfer_v.verify(t)$                       \COMMENT{verify proof} 
    \STATE   $ doTransfer(t.pub.pub)$                           \COMMENT{alg.~\ref{alg:trans}}
    \events        
\ENDINDENT
}

\subsubsection{Delegated Revealing Transfer Transaction}

The delegated revealing transfer is presented in algorithms~\ref{cir:del_rev} and~\ref{alg:del_rev}. The ZK circuit firstly execute the $checkDelegate$ function explained in Algorithm~\ref{cir:del_mint} and then the function $proveRevealingTransfer$ from Algorithm~\ref{cir:rev}. Finally, it converts the delegated revealing transfer witness (DRTW) to a proof.

\algoritmo{Delegated Revealing Transfer's ZK Circuit}{\label{cir:del_rev}}{
\STATE $DRTW \is RTW$                   \COMMENT{delegated revealing transfer's witness}        \label{data:del_rev_witness}
\STATE $DRTPI:$                         \COMMENT{delegated revealing transfer's public inputs}  \label{data:del_rev_pub}
    \STATE \tab $pub: RTPI,$            \COMMENT{transfer's public inputs}                      \label{data:del_rev_pub.pub}
    \STATE \tab $del_{add}: address$    \COMMENT{delegated contract address}                    \label{data:del_rev_pub.del_add}  
    \STATE \tab $del_b: uint256$        \COMMENT{delegated contract binding}                    \label{data:del_rev_pub.del_bind}  
    \STATE \tab $proof: uint256$        \COMMENT{ZK proof}                                      \label{data:del_rev_pub.proof}  
\vspace{2.5mm}
\INDENT{\textbf{proveDelRevTransfer}$(wit: DRTW, pub: DRTPI): uint256$}
    \require $ checkDelegate(wit, pub)$                 \COMMENT{alg.~\ref{cir:del_mint}}
    \STATE   $ proveRevealingTransfer(wit, pub.pub)$    \COMMENT{alg.~\ref{cir:rev}}
    \RETURN  $ convertToProof(wit)$
\ENDINDENT
}

The smart contract received a delegated revealing transfer transaction (DRTT) and check whether the sender is the delegated and verify the proofs. After, it executes the transfer and emit the events.

\algoritmo{Delegated Revealing Transfer Transaction and Smart Contract}{\label{alg:del_rev}}{
\STATE $DRTT:$                          \COMMENT{delegated revealing transfer's transaction}    \label{data:del_rev}     
    \STATE \tab $pub: DRTPI,$           \COMMENT{delegated revealing transfer's public inputs}  \label{data:del_rev.pub}
    \STATE \tab $proof: uint256$        \COMMENT{ZKP}                                           \label{data:del_rev.proof}
\vspace{2.5mm}    
\INDENT{\textbf{+ delegatedRevealingTransfer}$(t: DRTT)$}
    \require $ msg.sender \eq t.del_{add}$                      \COMMENT{caller is the delegate}
    \require $ del\_rev\_transf_v.verify(t)$                    \COMMENT{verify proof} 
    \STATE   $ doRevealingTransfer(t.pub)$                      \COMMENT{alg.~\ref{alg:rev_trans}}
    \events        
\ENDINDENT
}


\subsubsection{Delegated Hiding Transfer Transaction}

The delegated hiding transfer is presented in algorithms~\ref{cir:del_hid} and~\ref{alg:del_hid}. 
The ZK circuit firstly execute the $proveRevealingTransfer$ function explained in Algorithm~\ref{cir:hid} and then verify if the delegated is correct. 
Finally, it converts the delegated hiding transfer witness ($DHTW$) to a proof.

\algoritmo{Delegated Hiding Transfer's ZK Circuit}{\label{cir:del_hid}}{
\STATE $DHTW \is HTW$                   \COMMENT{delegated hiding transfer's witness}       \label{data:del_hid_witness}
\STATE $DHTPI:$                         \COMMENT{delegated hiding transfer's public inputs} \label{data:del_hid_pub}
    \STATE \tab $pub: HTPI,$            \COMMENT{hiding transfer's public inputs}           \label{data:del_hid_pub.pub}
    \STATE \tab $del_{add}: address$    \COMMENT{delegated contract address}                \label{data:del_hid_pub.del_add}  
    \STATE \tab $del_b: uint256$        \COMMENT{delegated contract binding}                \label{data:del_hid_pub.del_bind}  
    \STATE \tab $proof: uint256$        \COMMENT{ZK proof}                                  \label{data:del_hid_pub.proof}  
\vspace{2.5mm}
\INDENT{\textbf{proveDelHidTransfer}$(wit: DHTW, pub: DHTPI): uint256$}
    \STATE   $ proveHidingTransfer(wit, pub.pub)$                           \COMMENT{alg.~\ref{cir:hid}}
    \STATE   $ hash_{wit} \get hash256(wit)$
    \require $ pub.del_b \eq hash256(pub.del_{add}, hash_{wit})$
    \RETURN  $ convertToProof(wit)$
\ENDINDENT
}

The smart contract received a delegated hiding transfer transaction ($DHTT$) and check whether the sender is the delegate, then it verifies the proofs. 
Finally, it executes the hiding transfer and emit the events.

\algoritmo{Delegated Hiding Transfer's Smart Contract Flow}{\label{alg:del_hid}}{
\STATE $DHTT:$                        \COMMENT{delegated hiding transfer's transaction}     \label{data:del_hid}     
    \STATE \tab $pub: DHTPI,$         \COMMENT{delegated hiding transfer's public inputs}   \label{data:del_hid.pub}
    \STATE \tab $proof: uint256$      \COMMENT{ZKP}                                         \label{data:del_hid.proof}
\vspace{2.5mm}    
\INDENT{\textbf{+ delegatedHidingTransfer}$(t: DHTT)$}
    \require $ msg.sender \eq t.del_{add}$               \COMMENT{caller is the delegate}
    \require $ del\_rev\_transf_v.verify(t)$             \COMMENT{verify proof} 
    \STATE   $ doHidingTransfer(t.pub)$                  \COMMENT{alg.~\ref{alg:hid_trans}}
    \events        
\ENDINDENT
}

\section{A Sample Business Case: DvP Smart Contract} \label{sec:usecase}

This article introduces a powerful recipe with a multitude of potential applications already envisioned. 
While subsequent parts of this series will delve into these diverse use cases in greater detail, we begin here with an initial, illustrative example: 
the Delivery versus Payment (DvP) scenario. 
This foundational use case serves as an introductory gateway, showcasing the recipe's core strengths and providing a glimpse into its broader capabilities.  
This exploration of DvP will lay the groundwork for understanding the more complex applications presented later, demonstrating how this recipe can revolutionize traditional processes and unlock new efficiencies.

Delivery versus Payment, or DvP, is a settlement principle in finance where the transfer of a security or asset occurs simultaneously with the corresponding payment, typically in cash. 
This seemingly simple concept is, in reality, a significant innovation enabled by blockchain technology and its ability to create a secure and near-instantaneous settlement environment. 
In a DvP transaction powered by blockchain, both legs of the transaction - the delivery of the asset and the payment - are linked together atomically, meaning either both actions happen at the same time, or neither does.

The core value proposition of DvP lies in its ability to eliminate principal risk, the risk that one counterpart defaults after the other has already fulfilled their obligation.  
Traditional settlement processes are often lengthy and complex, involving multiple intermediaries, creating a window of time where this risk is very real. 
Blockchain's decentralized and immutable ledger, coupled with smart contracts that automatically enforce the terms of the DvP agreement, drastically reduces this risk, leading to a more efficient, secure, and trustworthy financial ecosystem. 
This is particularly important for high-value transactions or those involving less liquid assets.

\begin{figure}[!ht]
    \centering
    \includegraphics[width=1\linewidth]{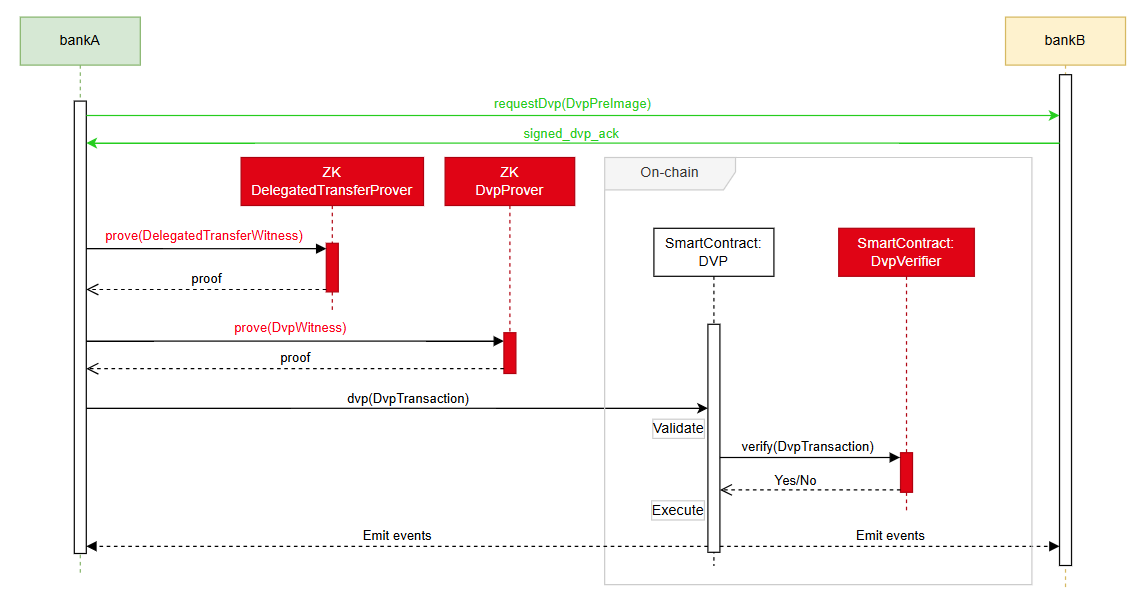}
    \caption{DvP start flow}
    \label{fig:dvp_start_flow}
\end{figure}

The diagram at Figure~\ref{fig:dvp_start_flow} illustrates a sequence of interactions between two banks, $bankA$ and $bankB$, engaging in the first part of a Delivery versus Payment (DvP) transaction facilitated by zero-knowledge (ZK) proofs and on-chain smart contracts. 
The process begins with $bankA$ initiating a request for DvP ($requestDvp$) to $bankB$, which responds with a signed acknowledgment ($signed\_dvp\_ack$). 
Subsequently, $bankA$ employs a $DelegatedTransferProver$ ZK circuit to generate a proof based on a $DelegatedTransferWitness$ ($DTW$) and its $DelegatedTransferPublicInputs$ ($DTPI$), and a $DvpProver$ ZK circuit to generate a proof using a $DvpWitness$ and its $DvpPublicInputs$.
These proofs are sent to the on-chain environment. 
The on-chain environment consists of two smart contracts: $DVP$, responsible for validating and executing the DvP transaction, and $DvpVerifier$, which verifies the ZK proof. 
Upon successful verification, the DvP smart contract executes the transaction. 
The completion of the process is indicated by both smart contracts emitting events. 

\data{DvP preimage}{\label{alg:dvp_pre}}{
\STATE $DvP preimage:$                                  \COMMENT{DVP preimage}                     \label{data:dvp_pre_img}
    \STATE \tab $[inputs: List\langle NPre \rangle,]$    \COMMENT{optional, nullifiers' preimages}  \label{data:dvp_pre.nullifier}
    \STATE \tab $outputs: List\langle TPre \rangle,$     \COMMENT{transferred tokens' preimages}    \label{data:dvp_pre.output}
    \STATE \tab $delivery: List\langle TPre \rangle,$    \COMMENT{expected tokens preimages}        \label{data:dvp_pre.delivery}
}

\begin{figure}[!ht]
    \centering
    \includegraphics[width=1\linewidth]{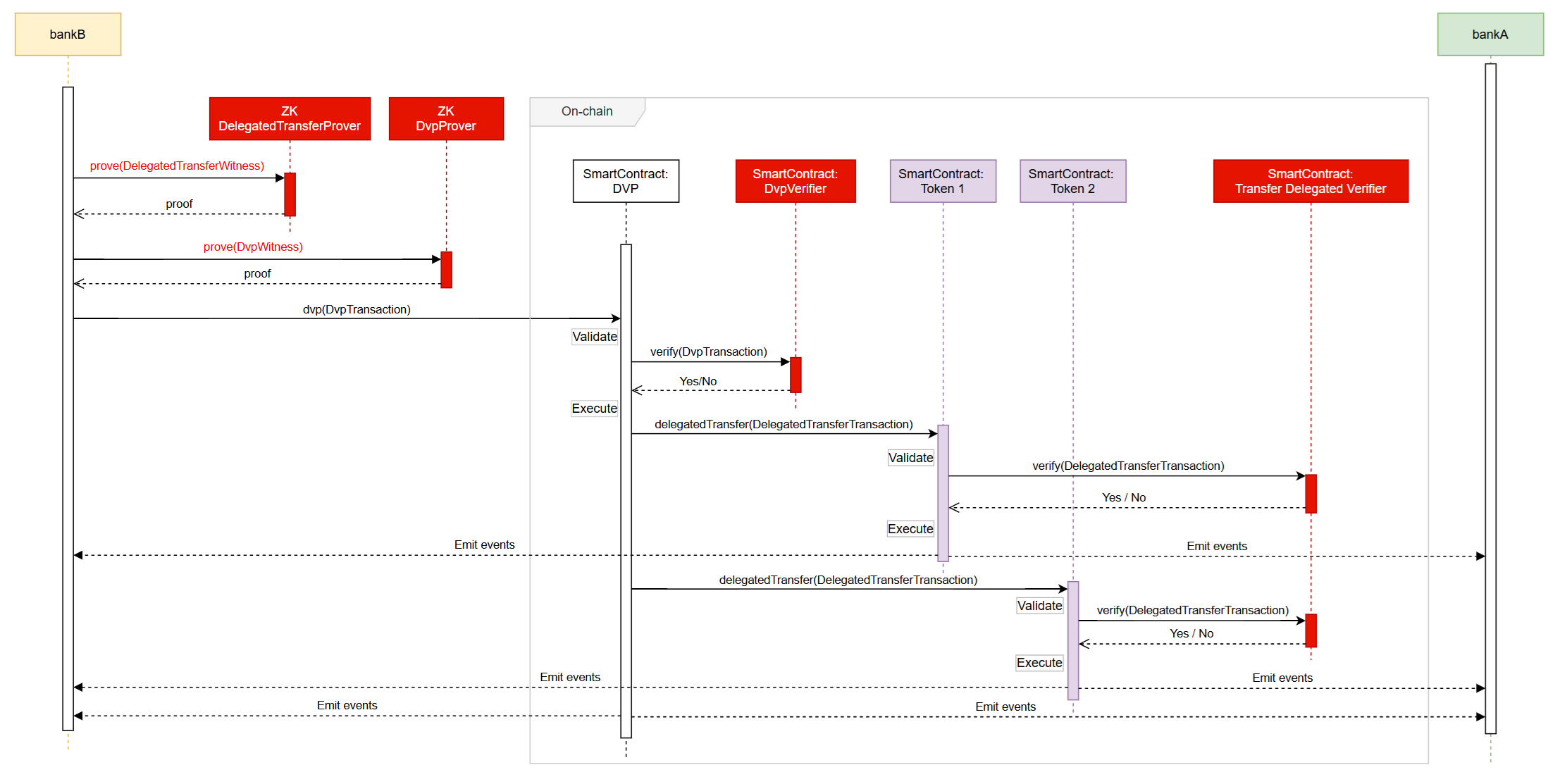}
    \caption{DvP confirmation flow}
    \label{fig:dvp_confirm_flow}
\end{figure}

The diagram in Figure~\ref{fig:dvp_confirm_flow} illustrates the concluding steps of a Delivery versus Payment (DvP) transaction involving two banks, $bankA$ and $bankB$, mediated by zero-knowledge (ZK) proofs and multiple on-chain smart contracts. 
Initially, $bankA$ utilizes a ZK $DelegatedTransferProver$ to generate a proof based on a $DelegatedTransferWitness$ ($DTW$) and its $DelegatedTransferPublicInputs$ ($DTPI$). Then a ZK $DvpProver$ to produce a proof using a $DvpWitness$ ($DVPW$) and the respective $DvPPublicInputs$ ($DVPPI$). 
These proofs are then submitted to the on-chain environment in a $DVPTransaction$ ($DVPT$). 
The on-chain infrastructure comprises five smart contracts: $DVP$, $DvpVerifier$, $Token1$, $Token2$, and $TransferDelegatedVerifier$. 
The $DVP$ contract initiates the validation process by invoking the $DvpVerifier$ to verify the ZK proof provided by the $DvpProver$ circuit, then it triggers the $delegatedTransfer$ function of token contracts. 
The $Token1$ and $Token2$ contracts, in turn, call upon the $TransferDelegatedVerifier$ to validate the delegated transfer transactions ZK proofs. 
Upon successful verification at each stage, the $Token1$ and $Token2$ contracts execute the fund transfers. 
The finalization of the process is signaled by each smart contract emitting events, which are observed by both $bankA$ and $bankB$, confirming the completion of the DvP transaction. 

The $DvP$ ZK circuit (Algorithm~\ref{cir:dvp}) works with its specific data, including a witness ($DvpW$) that contains the payment witness ($payment_w$) and a list of as tokens preimages ($TPre$) as its deliveries witness ($delivery_w$). 
It also uses public inputs ($DvpPI$), which include a full delegated transfer transaction ($payment$), a delivery hash ($delivery$), the delivery type ($type_d$), and a unique identifier that binds the DvP data together ($dvp\_bind$). 
The core function of the circuit is $proveDVP$, which generates a ZK proof. 
This function ensures that the $delivery$ in the public inputs is a hash of the $delivery\_w$ in the witness, and that the $dvp\_bind$ is a hash of the entire witness. 

\circuit{DvP}{\label{cir:dvp}}{
\INDENT{\textbf{Data Types:}}
    \STATE $(DvpW):$                                        \COMMENT{DVP's witness}                                     \label{data:dvp_witness}  
        \STATE \tab $payment_w: DTW,$                       \COMMENT{payment witness, alg.~\ref{cir:del_trans}}         \label{data:dvp_wit.payment}
        \STATE \tab $delivery_w: List\langle TPre \rangle$  \COMMENT{list of deliveries}                                \label{data:dvp_wit.delivery} 
    \STATE $DvpPI:$                                         \COMMENT{DVP's public inputs}                               \label{data:dvp_public_inputs}
        \STATE \tab $payment: DTT,$                         \COMMENT{alg.~\ref{alg:del_trans}}                          \label{data:dvp_pub.payment}  
        \STATE \tab $delivery: uint256,$                    \COMMENT{delivery hash}                                     \label{data:dvp_pub.d_hash} 
        \STATE \tab $type_d: uint256,$                      \COMMENT{delivery type}                                     \label{data:dvp_pub.d_type} 
        \STATE \tab $dvp\_bind: uint256$                    
\ENDINDENT
\vspace{2.5mm}
\INDENT{\textbf{proveDVP}$(wit: DvpW, pub: DvpPI) : uint256$}     
    \require $ pub.delivery \eq hash256(wit.delivery_w)$   
    \require $ pub.dvp_bind \eq hash256(wit)$
    \RETURN  $ convertoToProof(wit)$ 
\ENDINDENT
}

The $DvP$ smart contract acts as the enforcer of the DvP transaction rules on the blockchain. 
It corresponds to the $DVP$ smart contract shown in the Figures~\ref{fig:dvp_start_flow} and~\ref{fig:dvp_confirm_flow}. 
This smart contract uses a transaction data structure ($DvpT$) that includes the DvP's public inputs ($pub$) and the ZK proof ($proof$). 
The contract maintains a state that includes a reference to a verifier smart contract ($dvp_v$), a record of pending transactions ($pending$), and a list of addresses for token contracts ($contracts$). 
The smart contract's main function ($dvp$), processes incoming DvP transactions. 
It first verifies the DvP's ZK proof using the $dvp\_v$ contract and checks if it knows the token contract associated with the delivery type. 
It then checks for a matching pending transaction. 
If a match exists, it retrieves the relevant token contracts, $Token1$ and $Token2$, and calls the $delegatedTransfer$ function on both to transfer assets, finalizing the DvP process. 
If no match is found, the transaction is stored as pending. 
The contract emits events to signal when a transaction is completed or stored.

In summary, these algorithms describe how $bankA$ can use ZK proofs to demonstrate the correctness of a DvP transaction to the blockchain. 
The $DVP$ smart contract on the blockchain then verifies these proofs and coordinates with other smart contracts, including $Token1$, $Token2$, and $DelegatedTransferVerifier$, as shown in the Figure~\ref{fig:dvp_confirm_flow}, to ensure the atomic and secure settlement of the transaction between $bankA$ and $bankB$. 
The system ensures that asset transfers happen only if the corresponding delivery has been made, all while maintaining the privacy and integrity of the transaction details.

\contract{DvP}{\label{alg:dvp}}{
\INDENT{\textbf{Data Types:}}   
    \STATE $DvpT:$                       \COMMENT{DvP's transaction}    \label{data:dvpt} 
        \STATE \tab $pub: DvpPI,$        \COMMENT{DvP's public inputs}  \label{data:dvpt.pub}
        \STATE \tab $proof: uint256$     \COMMENT{ZKP}                  \label{data:dvpt.proof}
\ENDINDENT
\vspace{2.5mm}    
\INDENT{\textbf{State Variables:}}  
    \STATE $ dvp_v: DvpVerifier$                                \COMMENT{verifier Smart Contract}  \label{var:dvp.verifier}         
    \STATE $ pending: Map\langle uint256,DvpT \rangle $         \COMMENT{pending transactions}     \label{var:dvp.pending}         
    \STATE $ contracts: Map\langle uint256, address \rangle $   \COMMENT{token contracts}          \label{var:dvp.contracts}                 
\ENDINDENT
\vspace{2.5mm}    
\INDENT{\textbf{+ dvp}$(t_1: DvpT)$}                                                        \label{alg.dvp}
    \require $ this.dvp_v.verify(t_1)$                          \COMMENT{verify the ZKP}    \label{alg.dvp.req.verify}
    \require $ this.contracts[t_1.pub.type_d] \ne \bot$         \COMMENT{know the token}  
    \STATE   $ t_2 \get this.pending[t_1.pub.delivery]$
    \IF {$ t_2 \ne \bot$}
        \STATE $ contract_{Token1} \get contracts[t_2.pub.type_d]$
        \STATE $ contract_{Token2} \get contracts[t_1.pub.type_d]$
        \STATE $ contract_{Token1}.delegatedTransfer(t_1.pub.payment)$
        \STATE $ contract_{Token2}.delegatedTransfer(t_2.pub.payment)$        
        \STATE $ pending[t_1.pub.delivery] \get \bot$
    \ELSE
        \STATE $ pending[t_1.pub.delivery] \get t_1$       \COMMENT{store as pending}
    \ENDIF
    \events
\ENDINDENT
}

\section{Conclusions}
\label{sec:conclusion}

This paper presented a novel solution leveraging zk-SNARKs to enhance privacy in smart contracts and blockchain transactions. 
Our approach overcomes existing limitations, supporting both fungible and nonfungible tokens.
By enabling secure, decentralized, and private transactions, our solution facilitates broader blockchain adoption. 
The proposed delegated transaction mechanism expands use cases, such as Delivery vs Payment (DvP). 
Our findings demonstrate the feasibility of developing privacy in blockchain technology, paving the way for future research and real-world applications.

\bibliographystyle{IEEEtran}
%
\bibliography{references}

\end{document}